\documentclass[preprint,12pt]{elsarticle}

\usepackage{amsmath,amssymb}
\usepackage{lipsum}
\usepackage{outlines}
\usepackage{multirow}
\usepackage{float}

\usepackage{lineno}
\usepackage{etoolbox,refcount}
\usepackage{multicol}
\usepackage{amsmath,amsthm,amsfonts,amssymb,amscd}
\usepackage{hyperref}
\usepackage{natbib}
\usepackage{cleveref}
\usepackage{rotating}
\usepackage{tabularx}
\usepackage{protosem}
\usepackage{color}
\usepackage{xcolor}
\usepackage[scr]{rsfso}
\usepackage{natbib}
\usepackage{siunitx}

\journal{Journal of Fusion Energy}

\begin{document}

\begin{frontmatter}



\title{Lithium Experimental Application Platform (LEAP): Secondary-Containment Architecture for Flowing Liquid Lithium in Fusion Systems}


\author[PPPL]{Yufan Xu}
\author[ANL,MSU]{Yoichi Momozaki}
\author[PPPL]{Michael Hvasta}
\author[PPPL]{Robert Kaita}
\author[PPPL,PU]{Egemen Kolemen}

\affiliation[PPPL]{organization={Princeton Plasma Physics Laboratory}, 
            addressline={100 Stellarator Rd}, 
            city={Princeton},
            postcode={08540}, 
            state={NJ},
            country={USA}}
            
\affiliation[ANL]{organization={Argonne National Laboratory},
            addressline={9700 S Cass Ave}, 
            city={Lemont},
            postcode={60439}, 
            state={IL},
            country={USA}}

\affiliation[MSU]{organization={Michigan State University},
            addressline={640 S Shaw Ln}, 
            city={East Lansing},
            postcode={48824}, 
            state={MI},
            country={USA}}

\affiliation[PU]{organization = {Princeton University},
            addressline = {Engineering Quadrangle},
            city = {Princeton},
            postcode = {08540},
            state = {NJ},
            country = {USA}}


\begin{abstract}
Flowing liquid lithium is a promising fusion technology because it can provide a renewable Plasma-Facing Component (PFC) surface, modify recycling, support power exhaust, and potentially connect plasma-facing components with fuel recovery. Its deployment, however, is limited by the need to manage chemical reactivity, fire and aerosol hazards, inert gas operation, maintainability, and rapid experimental iteration. This paper develops a semi-quantitative hazard complexity framework for selecting secondary containment architectures for flowing liquid lithium systems. The framework is applied to six representative containment scenarios and to the Lithium Experimental Application Platform (LEAP) at Princeton Plasma Physics Laboratory. LEAP is under construction with a modular, room-scale argon gloveroom as an inert secondary containment boundary for a staged flowing lithium program with heating, diagnostics, magnetic field exposure, and future device interface capability. The analysis shows that an inert, airtight secondary enclosure without scrubbers around a liquid lithium loop provides a practical balance between hazard reduction and facility complexity, as defined by the design requirements. The resulting architecture offers a deployable path for lithium PFC development and a transferable design logic for other reactive or conductive liquid metal systems.
\end{abstract}





\end{frontmatter}



\section{Introduction}
\label{introduction}

\subsection{Liquid lithium systems for fusion reactors}

Liquid metals have high thermal and electrical conductivity and unique surface properties, making them desirable candidates for a wide range of applications in science and engineering (e.g., \cite{muller2001magnetofluiddynamics, davidson2017introduction, tang2021gallium}). Liquid lithium, the lightest alkali metal ($Z=3$), and lithium alloys are especially attractive for fusion science and technology because they have been proposed to address three key challenges at once: (i) self-replenishing plasma-facing components (PFCs) that survive extreme heat and particle fluxes, (ii) improved confinement via modified boundary conditions, and (iii) breeding, retention, transport, and recovery of fusion fuel species. There has been a decades-long interest in implementing liquid-metal plasma-facing components and liquid-metal fuel-cycle concepts for fusion devices \cite{coenen2014liquid, ono2017liquid, kaita2019fusion, de2021lithium, urgorri2018magnetohydrodynamic, doe2025fusionroadmap} since the pioneering work by Golubchikov \textit{et al.} (1996) \cite{golubchikov1996development} and subsequent experiments on Russian T-11 tokamak using liquid lithium capillary porous system (CPS) \cite{mirnov2003li}. 


Flowing lithium used as a plasma-facing component offers a self-renewing interface: the exposed surface can be continuously replenished, which reduces cumulative damage from intense heat and particle fluxes. In contrast, a solid plasma-facing component presents a stationary near-surface region, and heat removal is ultimately constrained by conduction through the solid and by allowable surface temperatures set by melting, recrystallization, and thermomechanical limits. Without resurfacing, solid PFCs can also undergo self-amplifying degradation: leading edges, gaps, and tile/module misalignment concentrate the parallel heat flux and create localized hot spots. These overloads can drive cracking, recrystallization, and melting in solid divertor components, while the resulting surface deformation further worsens nonuniformity in heat flux \cite{herrmann2015solid,nygren2017thermal,pitts2017physics}. Transient loads can also trigger crack growth and melt-layer motion or erosion, accelerating damage once overheating begins \cite{sinclair2017melt}. Arcing provides another localized erosion mechanism and can become more likely on damaged metal surfaces \cite{wang2020thermal}. 

Unlike solid materials, liquid lithium can leverage forced convection as an added heat-removal channel, so heat-flux handling can be controlled by tailoring flow velocity, wetted geometry, and heat extraction. Several concept families have therefore been pursued, including divertorlets \cite{fisher2020liquid,saenz2022divertorlets}, thin-film fast flows \cite{morley1995modeling,pan2025magnetohydrodynamic,jiang2026design}, capillary-porous systems with flow (CPSF) \cite{khodak2022plasma}, LiMIT-style thermoelectric-MHD-driven surfaces \cite{ruzic2011lithium}, and lithium vapor-box divertors \cite{goldston2016lithium,emdee2019simplified}. These concepts span a broad design space for lithium-enabled power exhaust, including low-recycling flowing-surface approaches as well as vapor-shielded, high-recycling divertor regimes. Experiments and simulations have demonstrated liquid-lithium PFC prototypes operating at incident heat fluxes of order $\sim \SI{10}{MW/m^2}$ \cite{ruzic2017flowing,saenz2022divertorlets,islam2024analysis,jiang2026design} while avoiding excessive evaporation. Although so far the heat handling capabilities of these prototypes remain below the $\mathcal{O}(10^2)~\SI{}{MW/m^2}$ levels often projected for divertors in compact pilot-plant or reactor concepts (e.g., \cite{kuang2020divertor}), flowing lithium remains an attractive research direction. Ongoing experiments and near-term demonstrations are expected to clarify the achievable operating window and the key physics and engineering constraints that will determine its viability for reactor-relevant power exhaust \cite{coburn2026overview}.

High-speed flowing lithium systems developed outside fusion have already demonstrated engineering reliability and stability in vacuum systems. In accelerator applications, liquid lithium has been engineered as a self-renewing charge-stripping medium for high-power heavy-ion linear accelerators (\cite{nolen2001argonne}). Conventionally, a thin carbon charge stripper is often installed in a linear particle accelerator to increase the beam charge state by removing electrons. However, these short-lived carbon strippers suffer from sublimation and radiation damage, thereby limiting the beam power density. The Facility for Rare Isotope Beams (FRIB) at Michigan State University (MSU) \cite{wei2019advances} established a flowing lithium film-formation method in which a round lithium jet impinges on a polished deflector to form a stable thin film and demonstrated high-vacuum operation with a prototypical high-pressure lithium system that produced a $\sim \SI{9}{mm}$-wide film at $\sim \SI{58}{m/s}$ with an estimated thickness $\lesssim \SI{13}{\mu m}$. The lithium stripper system withstands an impressive volumetric heat load from the proton beam, estimated at a maximum of $\SI{65}{MW/cm^3}$. As a result, heavy-ion stripping performance has also been successfully demonstrated while avoiding solid-foil lifetime limits \cite{momozaki2015proton, kanemura2022experimental}. Although an accelerator stripper is not directly equivalent to a fusion first wall, it demonstrates that engineered lithium flows can already be formed, stabilized, and operated in a demanding high-vacuum, high-thermal-load environment. This result could be particularly relevant to non-magnetic fusion concepts, where flowing-lithium plasma-facing surfaces may benefit from lithium’s self-renewing and heat-handling properties without incurring the full penalty of strong MHD constraints. Moreover, other lithium jet experiments have already established assessment and practice for handling lithium \cite{furukawa2014current,kondo2009liquid,knaster2016assessment,knaster2017assessment}.

\begin{figure}[ht!]
    \centering
    \includegraphics[width=\linewidth]{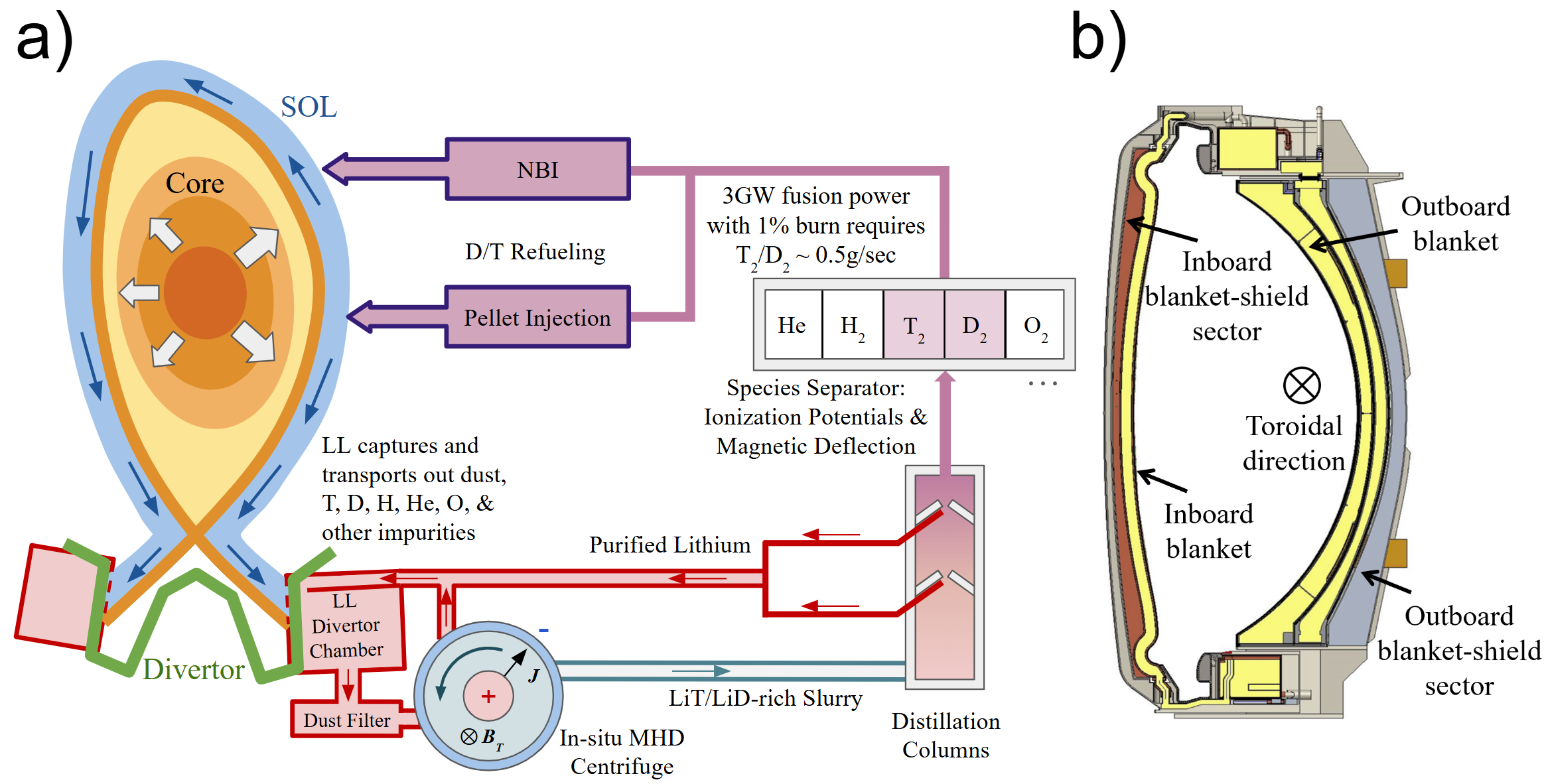}
    \caption{Lithium systems in a fusion reactor. a) Schematics of a liquid lithium divertor with D/T extraction at the first wall. D/T enters the lithium divertor via a low-recycling regime, forming lithium deuteride, LiD, and lithium tritide, LiT. These lithium hydrides are further concentrated via an in-situ MHD centrifuge and transported to a distillation column system, where deuterium and tritium can be extracted and used as fuels. Figure adapted from Ono \textit{et al.}, Nucl. Fusion (2017) \cite{ono2017liquid}, and de Castro \textit{et al.}, Phys. Plasmas (2021) \cite{de2021lithium} b) Example of a fusion breeding blanket design. Figure adapted from the lead-lithium blanket design study of Spherical Tokamak Advanced Reactor (STAR) \cite{gupta2026mhd}. }
    \label{fig:LowRecycle}
\end{figure}
Lithium’s affinity for hydrogen isotopes lowers wall recycling, which can increase edge neutrals temperature, relax edge temperature gradients, and reduce the turbulence drive associated with temperature gradients from cold neutral returning from the wall \cite{majeski2006enhanced}. This lithium-accessible low-recycling regime paved a new way for future fusion reactor designs and operations. Across devices such as CDX-U \cite{majeski2006enhanced}, LTX-$\beta$\cite{boyle2023extending}, NSTX/NSTX-U \cite{kugel2012nstx}, FTU \cite{mazzitelli2011ftu}, TJ-II \cite{tabares2008plasma}, HT-7 and EAST \cite{zuo2011first}, lithium conditioning or lithium-based PFCs, such as limiters, have been associated with improved confinement, modified edge profiles, and, in some operating regimes, mitigation or suppression of ELM activity \cite{maingi2009edge,zuo2013lithium}.

The deuterium/tritium (D/T) fuel cycle functions as the ``circulatory system'' of a fusion plant \cite{abdou2021physics}. Tritium must be bred, extracted, processed, delivered to the plasma, recovered from exhaust streams, and continuously recycled while maintaining rigorous accountancy and minimizing permeation losses. In most pilot-plant and reactor concepts, lithium-bearing blankets are the primary pathway to tritium self-sufficiency \cite{sawan2006physics, gupta2026mhd, baus2023kyoto}, which tightly couples blanket materials, tritium transport, and fuel-processing performance. In addition, tritium trapped in the first wall and divertor must be actively recovered and routed back into the fuel cycle, motivating flowing-liquid architectures that can provide both a low-recycling boundary and a controllable extraction pathway. 

A further coupling arises because a lithium plasma-facing layer can itself become a significant tritium breeding region. In a three-dimensional neutronics study of an FNSF-like design with a $\SI{2.51}{cm}$ liquid-metal first wall, Bohm \textit{et al.}~(2019) \cite{bohm2019initial} found that a lithium first wall increased the Tritium Breeding Ratio (TBR) by approximately $13\%$ relative to the baseline when tritium produced in the lithium first wall was assumed to be recovered, but reduced the effective TBR by approximately $31\%$ when that tritium was not recovered \cite{bohm2019initial}. In that study, the lithium also reduced peak neutron-induced damage to the first wall and hydrogen/helium production relative to the baseline, indicating potential neutronics benefits beyond plasma-surface renewal. Nevertheless, the large sensitivity of TBR to first-wall tritium recovery means that reactor-relevant lithium PFCs must be treated as part of the fuel-cycle system, not only as heat-exhaust or wall-conditioning components. This strengthens the need for flowing-lithium test stands that can investigate intermediate questions of hydrogen-isotope transport, concentration, extraction, and recirculation in addition to MHD flow and plasma-facing performance.

\Cref{fig:LowRecycle} a) illustrates a conceptual liquid-lithium divertor in which D/T species entering the divertor are absorbed into the lithium and converted into lithium hydrides, which are then removed from the plasma-facing region, concentrated (e.g., by an in-line MHD-based centrifuge), and processed in downstream systems such as distillation column systems to recover and re-inject deuterium and tritium into the fueling loop. \Cref{fig:LowRecycle} b) shows a representative lead-lithium breeding blanket configuration in the design study of the Spherical Tokamak Advanced Reactor (STAR), highlighting that the blanket design, tritium breeding and extraction, and primary heat removal are inherently coupled at the system level.

Accordingly, the central question for liquid lithium systems in fusion is no longer whether they are scientifically compelling, but whether the remaining science and engineering gaps can be closed on timelines compatible with reactor and pilot-plant deployment, and whether they can be effectively integrated with existing fusion pilot-plant designs. The DOE Fusion Science \& Technology Roadmap \cite{doe2025fusionroadmap} highlights several limiting gaps, including plasma shielding and contamination control; free-surface behavior and MHD flow stability; loop-scale heat transfer and component reliability; corrosion and compatibility of lithium-exposed materials; impurity accumulation and clogging; practical D/T extraction, inventory control, and recirculation; and the shortage of integrated, design-quality datasets that couple neutronics, thermofluids, and MHD. These needs motivate dedicated lithium and lithium-alloy test stands that operate at relevant temperature, flow, and magnetic conditions to benchmark tritium transport and extraction, quantify corrosion and impurity effects, validate MHD-influenced thermohydraulic behavior, and demonstrate integration with tritium processing subsystems. Consistent with the Roadmap facility strategy, such platforms should mature from component-level experiments to integrated blanket and fuel-cycle testing, providing validated operating envelopes and an engineering basis for pilot-plant design and licensing.

\subsection{Legacy experience from alkali-metal fission systems}

Large-scale liquid alkali-metal infrastructures have been developed primarily for liquid-metal-cooled nuclear fission reactor facilities \cite{lane1958fluid}. The world’s first breeder reactor prototype, the Experimental Breeder Reactor-I (EBR-I), was commissioned in 1951 \cite{michal2001fifty} and used a sodium–potassium eutectic alloy (NaK) as the coolant. More recently, the Sodium-cooled Fast Reactor (SFR) has been pursued as a Generation IV reactor concept currently under development \cite{kultgen2022mechanisms}. Alkali-metal coolants offer several advantages over current water-based reactors. Because metal atoms are weak neutron moderators, fast-spectrum operation is facilitated, thereby enabling efficient transmutation and significantly reducing the production of transuranic waste \cite{gif_sfr_portal}. The high boiling points of sodium and NaK yield large available temperature margins, increasing the permissible temperature rise in the coolant. Consequently, liquid-metal-cooled systems can achieve higher thermodynamic efficiency than water-cooled systems, despite the lower specific heat capacities of liquid metals compared to water \cite{marshall2002_sand2002_0513}. Sodium and NaK are also relatively abundant and exhibit favorable chemical compatibility with structural steels and many types of nuclear fuel cladding materials \cite{jackson1955_liquid_metals_handbook_sodium_nak_supp}. Nevertheless, the deployment of large-scale alkali-metal coolant systems presents significant challenges arising from the corrosive and highly reactive nature of these metals, as well as their propensity to leak. Therefore, rigorous design measures and specialized operational protocols are required to prevent, detect, and mitigate fires and other hazardous events \cite{foust1978_sodium_nak_handbook_v3}. This need is underscored by several historical incidents involving sodium fires, such as the sodium leak and subsequent fire at the Monju Nuclear Power Plant in 1995 \cite{miyakawa2000sodium}. Classical sodium and NaK handbooks \cite{jackson1955_liquid_metals_handbook_sodium_nak_supp,foust1978_sodium_nak_handbook_v3} provide valuable guidance for components, handling, and safety practice. International efforts to build fission systems cooled by liquid metals or molten salts have also invested extensively in safety and liquid-metal materials compatibility, driven in large part by the requirements of liquid-metal fast breeder reactors (LMFBRs), e.g., \cite{hvasta2024mechanisms}. These lessons translate directly to fusion lithium facilities, which share many material and system-scale hazards. 

\subsection{Lithium-specific Safety Requirements}
Development and implementation of liquid lithium systems at a reactor scale remains challenging. Lithium characteristics and risks have been discussed extensively in previous studies \cite{de2021lithium,d2023lithium}. Liquid lithium produces exothermic reactions with air and moisture. Lithium's reaction with water generates hydrogen, creating explosion risk. Lithium is also considered pyrophoric in powder form or liquid at elevated temperatures. The flowing lithium systems must address general alkali-metal safety requirements, such as leak prevention, inerting strategy for exclusion of air and moisture, hydrogen-generation risk, fire consequence management, contamination control, maintainability, and reliable operation during off-normal events. Moreover, any leak or spill can generate corrosive and toxic aerosols and can mobilize radionuclides in tritium-containing lithium systems \cite{d2023lithium}. Past lithium loop incidents (e.g., \cite{maroni1981analysis,nygren2012pmtf_li_fire}) have already demonstrated credible leak and fire scenarios. 

As a result, the design of liquid lithium systems should mainly focus on its containment strategy and minimizing the contact of non-compatible or reactive materials with lithium. Using IFMIF-DONES as a concrete design study, D’Ovidio \textit{et al.} (2023) \cite{d2023lithium} presents a defense-in-depth fire protection strategy for large liquid-lithium facilities. A large-scale lithium system should implement robust primary confinement by mechanical codes and standards, minimization and segmentation of inventory where possible, room inertization (He/Ar) with gas monitoring, stainless-steel liners to prevent Li–concrete reactions, catch pans and drain and recovery systems, leak/smoke/flame detection, and compartmentalization isolation. A key conclusion is that for large lithium fires, fixed chemical-agent suppression may be unreliable, so passive measures (such as inerted rooms, spill capture and drainback, structural heat sinking, compartmentalization, etc.) are favored as the main line of defense, complemented by appropriately selected Class-D agents (e.g., Lith-X, Natrex-L, copper powder, carbon-based agents) and aerosol filtration strategies. Application of the MELCOR-fusion code has been demonstrated to analyze the transient accident associated with the lithium system for IFMIF-DONES test cell \cite{perez2023application}. 


A series of regulatory standards for alkali-metal systems has been established. The International Building Code (section 307, 2021 edition) defines the maximum allowable quantities for a lithium system in a controlled area \cite{IBC-2021-Section307}. For example, the code limits storage and closed-container use to $\SI{22.68}{kg}$ of lithium per control area (this increases to $\SI{45.36}{kg}$ per control area when lithium is stored or used in a ventilated enclosure). The limit is $\SI{4.54}{kg}$ for open systems. A building may be permitted to have multiple control areas, up to four in total.

On fire protection, U.S. National Fire Protection Association Standard for Combustible Metals (NFPA 484) \cite{NFPA484-2022} covers facility design requirements, fire and explosion protection, powder, processing, machining, fabricating, finishing, storage, and handling, etc., with applicable threshold masses. Other alkali metals fire protection standards can be found in The International Fire Code (IFC) \cite{IFC-2024}, and the related defense-in-depth framework is covered in Fire Protection in Nuclear Power Plants (IAEA-TECDOC-1944) by the International Atomic Energy Agency (IAEA) \cite{IAEA-TECDOC-1944}. It is strongly recommended, and in some cases required when the applicable NFPA 484 thresholds are met, to develop a fire protection plan and notify the local fire department, and make sure all personnel are trained to fight lithium fires.   

Plumbing and piping system design under pressure and heating can refer to ASME B31.3, Process Piping Code \cite{ASME-B31.3-2024}, and ASME BPVC, the Boiler and Pressure Vessel Code \cite{ASME-BPVC-2025}, if applicable. 

\subsection{Risk-informed approach with infrastructure and operational complexity}

It is sometimes neglected that safety systems that impose excessive procedural burden can increase operating cost and degrade overall safety by increasing the likelihood of human error. As outlined in human reliability analysis frameworks that have been used in nuclear safety, task complexity and procedure quality are performance-shaping factors that increase estimated human error probability. As a consequence, designs that require extensive and highly complex operational procedures tend to be more error-prone \cite{gertman2005spar}. 


These considerations motivate a fusion-oriented design framework that balances lithium safety requirements against infrastructure and operational complexity, enabling rapidly deployable research-scale flowing-lithium test stands. This need is most acute for platforms built for rapid iteration, where frequent hardware changes, evaluation of multiple PFC concepts, open-surface operation, strong magnetic fields, and eventual integration with a confinement device must all be accommodated on compressed development timelines. An integrated lithium platform, as a part of the technology development itself, should provide a stable safety boundary and standardized services that enable repeated test $\rightarrow$ modify $\rightarrow$ downselect cycles, while initiating a credible pathway toward integration of lithium systems and reactors.

In this study, we introduce a generalized design framework and secondary-containment choices for flowing liquid-lithium systems in fusion applications, with emphasis on safety, maintainability, and implementation in laboratory and industrial environments. Specifically, we present a reusable, risk-informed system engineering framework for selecting secondary-containment architectures for flowing liquid-lithium facilities and demonstrate the approach through the modular ``gloveroom'' concept used in the Lithium Experimental Application Platform (LEAP) at the Princeton Plasma Physics Laboratory (PPPL). The framework links mission requirements, including device-facing constraints, maintainability, prototyping flexibility, and facility readiness, to industrial safety and regulatory considerations. Although the present work centers on lithium, the design logic is intended to provide transferable engineering principles for a broader class of liquid metals and conductive fluids.

The paper is organized as follows. \Cref{method} introduces the hazard-complexity evaluation framework for lithium systems. \Cref{senarios} identifies a general downselect list of six secondary containment options. \Cref{LEAP} presents the design requirements for flowing-lithium systems in LEAP and describes the selected modular secondary-containment architecture as an example. Section \ref{discussion} discusses LEAP's interfacing strategy with fusion devices and considerations on the limitations of the gloveroom.

\section{Methodology for determining containment architecture}
\label{method}

There is no single secondary-containment solution that is optimal for every flowing lithium system. The appropriate design depends on the intended lithium inventory, operating modes, expected failure consequences, required flexibility for upgrades, and the degree of integration with facility infrastructure. For fusion applications, this trade space is especially important because the same platform must often support lithium safety, open-surface or vacuum interfaces, magnetic compatibility, maintainability, and rapid experimental iteration, etc. 

To make these tradeoffs explicit, we formulate a streamlined downselecting evaluation metric that compares candidate secondary-containment architectures based on two competing objectives: hazard and facility complexity. In this framework, hazard is reduced by limiting the probability and consequence of lithium exposure to reactive environments, whereas complexity represents the added infrastructure, controls, and operational burden required to achieve that reduction. In general, increasing the number and sophistication of safety features and supporting infrastructures within a system tends to reduce the frequency and severity of hazardous events, while concurrently increasing the overall complexity of the facility. Vice versa, systems with fewer safety provisions are typically associated with a higher level of hazard but reduced facility complexity. However, as discussed in the previous section, we must account for the possibility that safety and facility complexity can be intertwined: a less complex system could reduce operational burden and improve safety. Further, we assume that the main cost of the system is driven by the complexity of the hardware/software and the system's operation. The goal of this exercise is therefore to find the optimal solution that minimizes hazards and facility complexity at a given cost.

\subsection{Design Penalty Index}

To compare candidate secondary-containment architectures on a common basis, we introduce a semi-quantitative \emph{design penalty index}, $\mathcal{Q}$, that balances hazard reduction against system (infrastructure and operational) complexity. The index is intended as a comparative engineering tool for downselection, rather than as a licensing metric. Lower values of $\mathcal{Q}$ correspond to more favorable overall designs. 

We define two main parameters, the intrinsic hazard and the complexity risks. The hazard index $\mathcal{H}$ represents the severity of the relevant lithium-specific failure consequences for a given system, such as reaction with air or moisture, aerosol generation, hydrogen production, radionuclide mobilization, and oxygen-deficiency hazards associated with inerting. The complexity index $\mathcal{C}$ represents the added facility and operational risk required by the containment concept, including enclosure requirements, atmosphere-control systems, detection systems, scrubbing or recovery systems, and the associated procedural and administrative burden. These indices represent risks that can be written as a sum of individual expectations based on Bernoulli's expected utility hypothesis for risk analysis \cite{bernoulli2011exposition},
\begin{equation}
    \mathcal{H} = \sum_i p_i^{(\mathcal{H})} h_i,
    \qquad
    \mathcal{C} = \sum_j p_j^{(\mathcal{C})} c_j,
    \label{eq:HC}
\end{equation}
where $h_i \in [0,5]$ represents the severity level or impact of each hazard mode in the lithium system, with 0 indicating negligible impact and 5 the highest impact and consequence. $c_j\in[0,5]$ indicates the complexity impact associated with each safety feature, with 0 indicating negligible impact and 5 indicating the largest contribution to system complexity. In general, the higher $c_j$ is, the more demanding, costly, and thus riskier it is to deploy a given feature. The corresponding weights $p_i^{(\mathcal{H})}$ and $p_j^{(\mathcal{C})}$ represent the probability of occurrence for each hazard mode and probability of implementing certain safety features, respectively. 

We can further develop this from simple superposition to capture more nonlinear and intricate connections between hazards and complexity risks. It can be postulated that:
\begin{itemize}
    \item Different designs are limited by various constraints and thus focus differently on hazards and complexity risks. 
    \item A safety feature may become less effective when applied to a more hazardous system or when embedded in a more complex facility with more failure modes, interfaces, and operational dependencies.
\end{itemize}

These two conjectures lead to three coefficients: the hazard coefficient, $\alpha$; the complexity coefficient $\beta$; and the mitigation effectiveness, $\gamma_{\mathrm{eff}}$. The hazard coefficient, $\alpha \in [0,1]$, and the complexity coefficient, $\beta \in [0,1]$, define the practical design focus on addressing hazard and complexity risks. When $\alpha = 0$, the system design ignores the occurrence of any safety hazards, whereas facility complexity is not constrained at $\beta = 0$ (e.g., infinite budget and perfect graduate students). It should be noted that $\alpha$ and $\beta$ should be determined relatively to reflect on the design interest, since $\mathcal {H}$ is not necessarily equal to $\mathcal {C}$. In addition, the mitigation effectiveness parameter $\gamma_{\mathrm{eff}} \in [0,1]$ represents how effectively the selected safety architecture reduces the hazard in practice, where $\gamma_{\mathrm{eff}} = 0$ corresponds to an ineffective safety feature and $\gamma_{\mathrm{eff}} = 1$ corresponds to ideal, perfect hazard mitigation.

The design penalty index can be estimated as a simple mathematical model to qualitatively capture the basic interplay between the risks,
\begin{equation}
    \mathcal{Q} = \alpha \mathcal{H} \left(1-\gamma_{\mathrm{eff}}\right) + \beta \mathcal{C},
    \label{eq:Q}
\end{equation}
where the first term represents the residual hazard penalty after accounting for the effectiveness of the safety architecture, while the second term represents the direct penalty associated with system complexity. This form preserves the desired monotonic behavior: $\mathcal{Q}$ increases with larger hazard and greater complexity, and decreases with enhanced mitigation effectiveness.

The definition of mitigation effectiveness is less certain. In most occasions, we can simply assume $\gamma = 1$ if mitigation effectiveness is included, or $0$ if it can be ignored. However, because practical effectiveness may degrade as the intrinsic hazard and system complexity increase, we can choose to further model an effective mitigation coefficient as a function of $\mathcal{H}$ and $ \mathcal{C}$,
\begin{equation}
    \gamma_{\mathrm{eff}} = \gamma \exp\left(-k_H \mathcal{H} - k_C \mathcal{C}\right),
    \label{eq:gammaeff}
\end{equation}
where $k_H > 0$ and $k_C > 0$ are sensitivity parameters selected by users. The exponential form ensures $0< \gamma_{\mathrm{eff}} \leq \gamma$. The nominal mitigation coefficient $\gamma$ can itself be constructed from the selected safety features. For example, a binary $x_m \in \{0,1\}$ denotes the presence or absence of mitigation feature $m$, such as inerted secondary containment, atmosphere monitoring, drainback, spill capture, or aerosol filtration, then
\begin{equation}
\gamma = 1 - \exp\left(-\sum_m \eta_m x_m\right),
\label{eq:gamma}
\end{equation}
where $\eta_m$ is the contribution of feature $m$ to the overall mitigation effectiveness. This representation ensures that $\gamma$ remains bounded between $0$ and $1$, so that $0 \leq \gamma_{\mathrm{eff}} \leq 1$, while also capturing diminishing returns as more safety features are added.

The optimal secondary-containment concept is then obtained by minimizing $\mathcal{Q}$ over the set of admissible design choices $\mathbf{x} \equiv (x_1, x_2, x_3, ...) \in X$,
\begin{equation}
    \min_{\mathbf{x} \in X} \mathcal{Q}(\mathbf{x}),
    \label{eq:opt}
\end{equation}
subject to any additional project-specific constraints, such as lithium inventory, open-surface operation, magnetic-field compatibility, vacuum interfaces, available building infrastructure, or code-related requirements. This optimization also allows investigation of the optimal continuous design features (e.g., argon purge rate, oxygen and moisture set point, spill-capture capacity, and redundancy level) to understand which features are most valuable when added safety begins to have diminishing returns and how strongly complexity penalizes further mitigation. A detailed derivation of the optimization problem in \cref{eq:opt} is shown in \ref{app1}.

The design penalty index provides a transparent basis for comparing containment strategies, identifying the dominant safety drivers, while balancing risk reduction against implementation burden. This risk-informed matrix can be adapted to different classes of alkali-metal systems simply by changing the modes, their impacts, and probabilities. In the present work, this theoretical framework is used as a structured downselection method for representative containment scenarios and to support the modular inerted secondary-containment architecture adopted for LEAP. 
\clearpage

\begin{table}[t]
\footnotesize
\centering
\begin{tabular}{l|l|l|l}
\hline \hline
$i$ & Hazards & $h_i$ & Justification \\
\hline
1 & $\mathrm{H_2}$ generation with bulk water & 5 & Large quantity of hydrogen gas can be \\
 &  &   & generated in a short time. Explosion risk. \\
2 & $\mathrm{H_2}$ generation with air moisture & 4 & Accumulation of hydrogen over time. \\
 &   &   & Explosion risk.\\
3 & Lithium Fire with $\mathrm{O_2}$ & 3 & Exothermal reaction. Oxide also reacts with \\ 
 &   &   & water to form hydroxide.\\
4 & Lithium Fire with $\mathrm{N_2}$ & 2 & Mild exothermal reaction. Can generate \\
 &   &   & toxic ammonia gas with moisture and $\mathrm{O_2}$.\\
5 & Lithium aerosol/smoke & 3 & Reduce visibility, toxic, radionuclides\\
 &   &   & mobilization. Indication of fire.\\
6 & Asphyxiation by inert gas & 1 & Inert containment hazard, can be mitigated\\
 &   &   & through monitoring and control, not common.\\
\hline \hline
\end{tabular}
\caption{List of hazards and their impact, $h_i$, in a generic lithium system. These hazards will be considered in the analysis of this study.}%
\label{tab:hazards}
\end{table}
\begin{table}[h]
\footnotesize
\centering
\begin{tabular}{l|l|l|l}
\hline \hline
$j$ & Safety Feature & $c_j$ & Justification \\
\hline
1 & Watertight 2nd containment & 2 & Relatively low complexity. Simple waterproof \\
 &   &   & material and design.\\
2 & Airtight 2nd containment & 3 & Medium complexity. Airtight seal or vacuum \\
 &   &   & system are often available off-the-shelf. \\
 &   &   &  Requires maintenance and leak check. \\
3 & Wet scrubber  & 5 & Very complex and less common. Difficult to \\
 &   &   & build and maintain. Needs dedicated water \\  
 &   &   & supply and waste drain.\\
4 & Dry scrubber ($\mathrm{O_2}/\mathrm{N_2}$)  & 4 & Requires maintenance. Need to replace\\
 &   &   & filter/scrubber and add exhaust. Expensive.\\
5 & Ventilation \& exhaust & 3 & Medium complexity. Emergency exhaust to \\
 &   &   & avoid aerosol and hydrogen accumulation.\\
6 & Dehumidifier & 2 & Commercial product available. \\
 &   &   & Requires additional maintenance.\\
7 & $\mathrm{H_2}$ detector & 1 & Low impact. Easy to deploy. \\
8 & $\mathrm{H_2O}/\mathrm{O_2}/\mathrm{N_2}$ analyzer & 1 & Low impact. Easy to deploy.\\
\hline \hline
\end{tabular}
\caption{List of safety features and their impact, $c_j$, on complexity in a generic lithium system. These safety features will be considered in the analysis of this study.}%
\label{tab:complexity}
\end{table}
\clearpage

\subsection{Simplified parameters}

The probability and severity of lithium-related hazards are difficult to quantify precisely and are strongly dependent on the specific system configuration. For an initial evaluation of secondary-containment concepts, we therefore adopt a simplified and conservative assumption: the primary containment is assumed to fail with a lithium breach rate of 100\% for all candidate designs. Under this condition, the primary containment itself is not used to distinguish between architectures. Instead, the comparison focuses on how the selected secondary-containment concept modifies the operator or facility exposure to the resulting hazards. For example, hydrogen generation from moisture is more likely if leaked lithium is exposed to ambient air than if the same leak occurs within an inert-gas-filled secondary enclosure, even when both systems use highly reliable primary containment. This simplification allows the design matrix to isolate the safety value of the secondary-containment architecture itself.

The impact of the hazards $h_i$, and the impact on the complexity risk $c_j$, are both assumed to be generic and independent of the design. For simplicity, discrete, non-zero integers are used so that $h_i,\ c_j \in \{1,2,3,4,5\}$ in an ascending order of impact. \Cref{tab:hazards} shows common lithium system related hazards with justification that have been included in the analysis for this study and \cref{tab:complexity} shows a list of common safety features used in lithium systems and their impact on complexity with justification that has been considered in this study. Furthermore, $p_i^{(\mathcal{H})}$ is defined as 
\begin{equation}
    p_i^{(\mathcal{H})}  = 
    \begin{cases}
        1 & \text{full exposure upon failure} \\
        0.5 & \text{possible exposure to hazard but under control} \\
        0 & \text{no exposure to hazard} \\
    \end{cases}
    \label{eq:piH}
\end{equation}
This assumes the scenario that primary containment has failed. Here, ``under control'' means any lithium escaped from the primary containment is under automatic facility control so no or little manual operation is required. Similiarly, 
\begin{equation}
    p_j^{(\mathcal{C})}  = 
    \begin{cases}
        1 & \text{if feature is needed} \\
        0.5 & \text{if feature is optional} \\
        0 & \text{if feature is not needed} \\
    \end{cases}
    \label{eq:pjC}
\end{equation}
Therefore, the probability of having a certain feature has been effectively translated to whether this feature is needed in the system. For instance, an essential equipment has a 100\% probability of showing up in the system.

\begin{figure}[ht]
    \centering
    \includegraphics[width=\linewidth]{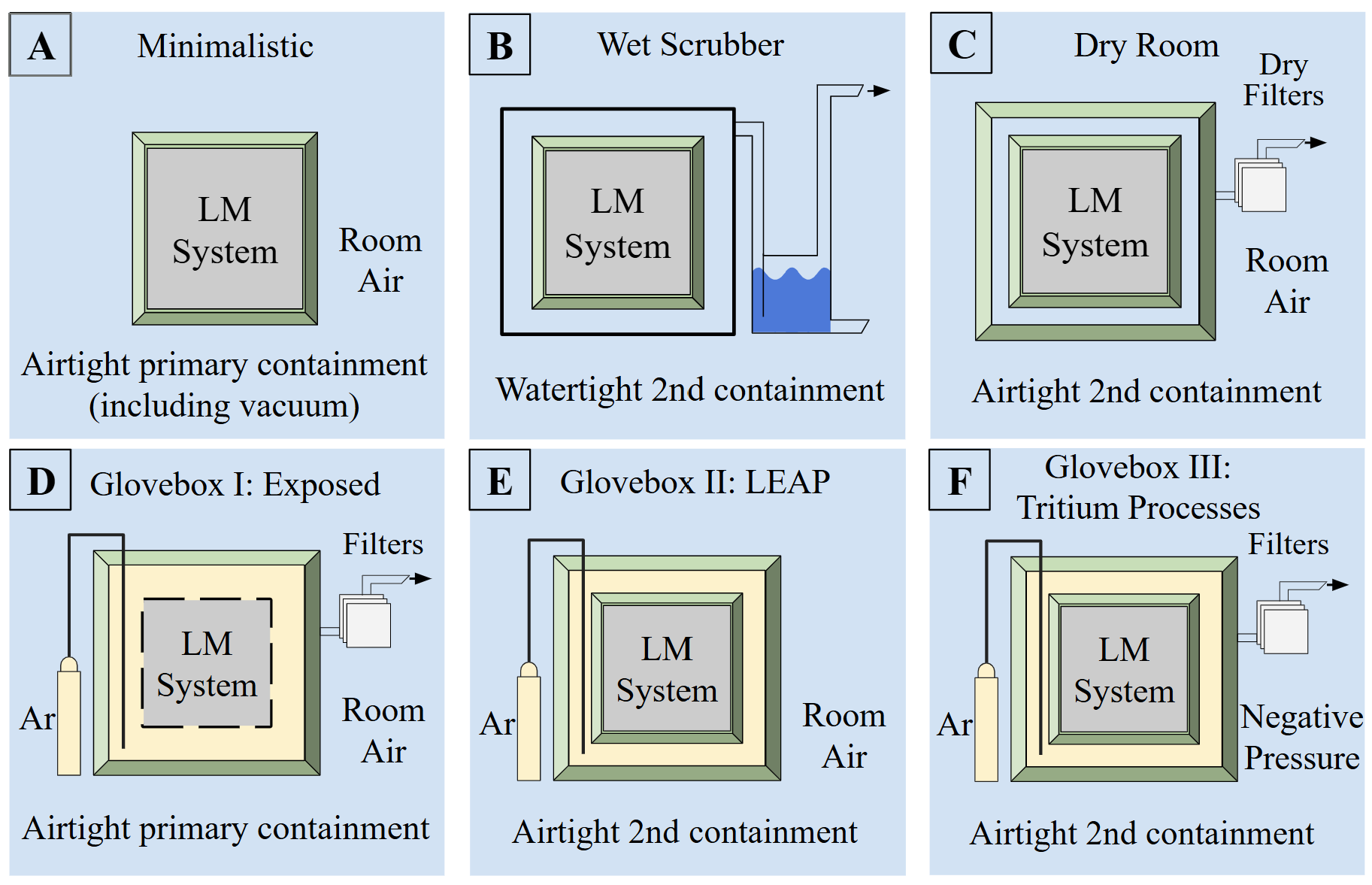}
    \caption{Six scenarios for the liquid metal containment systems. (A) A minimalistic setup: a primary containment for liquid lithium. (B) A watertight or semi-airtight secondary containment encloses the airtight primary containment and is also connected to a wet scrubber.  (C) A typical dry room: a humidity-controlled secondary containment is connected to a dehumidifier or a dry filter system. (D) A glovebox with open-surface liquid metal: airtight primary containment is filled with inert gas and connected to a dry filter system. (E) A basic glovebox setup: airtight secondary containment is filled and purged with inert gas, such as Argon, but does not connect to any filters or scrubbers. (F) A fully equipped glovebox with oxygen, humidity, and/or nitrogen filter and scrubbers. Negative pressure is implemented for containing radionuclides.}
    \label{fig:6scenarios}
\end{figure}

\section{Lithium system scenarios and requirements}
\label{senarios}

We can summarize existing lithium containment systems into six general example scenarios, as shown in the six panels of \cref{fig:6scenarios}. These six scenarios do not exhaust all the options for developing lithium systems. They are treated as representative cases that have been used in the downselect for this study. Each scenario corresponds to a unique set of $p_i^{(\mathcal H)}$ and $p_j^{(\mathcal C)}$. 


Scenario A (Minimalistic) setup is shown in \cref{fig:6scenarios} (A). Liquid lithium is contained inside an airtight primary containment, usually filled with inert gas, such as Argon, or under vacuum for free-surface operations. There is essentially only one barrier between liquid lithium and the room environment. This is the simplest approach for various tabletop lithium systems with a limited amount of lithium, and therefore has a low complexity risk. However, it is the least safe option because the leaked lithium would have direct exposure to the exterior environment with oxygen, nitrogen, moisture and even possibly bulk water in case of failure. \Cref{tab:p_iH} reflects that scenario A is under exposure to hydrogen generation via moisture, lithium fire, and smoke. Hydrogen generation with bulk water can be limited by facility controls, such as excluding nearby water sources or using dry or pre-action sprinkler systems. Nevertheless, scenario A needs the least amount of safety features (hydrogen detector and ventilation) and hence has the lowest complexity risk among all scenarios, as shown in \cref{tab:p_jC}.

Scenario B (Wet scrubber) shown in \cref{fig:6scenarios} (B) is a uniquely designed secondary containment (which can be an enclosure or a building) that connects to a wet scrubber system. In case of a liquid metal fire, the generated gas and smoke can be directed to a water container for reaction and passivation (e.g., the METL sodium system at Argonne National Laboratory). The wet scrubber does not completely mitigate the fire and hydrogen hazards, but has been implemented as a control method to constrain the consequences of a lithium incident. Scenario B in \cref{tab:p_iH} lists most of the hazards as under control, $p_i^{(\mathcal{H})} = 0.5$. Because there is a watertight secondary containment, it is unlikely lithium will see any bulk water to generate hydrogen, $p_1^{(\mathcal{H})} = 0$. However, wet scrubbers are less common in lithium processing, and they are difficult to build and maintain. They also require dedicated water supplies and waste drains that are isolated from lithium usage. Therefore, we consider that wet scrubber has the highest complexity risk impact among all the safety features (\cref{tab:complexity}). 

Scenario C (Dry room) shown in \cref{fig:6scenarios} (C) is a humidity-controlled only setup with air-tight secondary containment. These dry-room scenarios are commonly adopted in lithium battery production. Therefore, there is no moisture or bulk water, and hydrogen generation upon failure can be eliminated. Therefore, a hydrogen detector is optional. Lowering the moisture in the secondary containment also lowers the reaction speed of oxygen and nitrogen with lithium. However, exposed lithium can still ignite and react with oxygen and nitrogen in the air at a high temperature. Therefore, $p_i^{(\mathcal{H})} = 0.5$ for fire and smoke, as shown in \cref{tab:p_iH}.  

Scenario D (Glovebox I) shown in \cref{fig:6scenarios} (D) is a glovebox setup with dry filters and an open lithium surface. It is similar to Scenario A, except that an airtight containment is filled and purged with Argon gas. Argon purge and filtration will maintain low levels of oxygen, humidity, and nitrogen, often at $\sim 1$ ppm or lower, to minimize lithium reaction with air inside the glovebox. This glovebox is essentially just an atmosphere-controlled primary containment under near-room pressure. This is often considered safe for a small amount of lithium (grams) operation. However, due to the lack of secondary containment, $p_i^{(\mathcal{H})} = 0.5$ for all hazards. Because the enclosure is inerted with argon, oxygen monitoring is required to manage oxygen-deficiency hazards.

Scenario E (Glovebox II) shown in \cref{fig:6scenarios} (E) is a simple glovebox or enclosure with reduced functionalities: airtight secondary containment around an enclosed primary containment, and is filled and purged with inert gas, such as Argon. This setup does not connect to any filters or scrubbers, and generally can reach $\lesssim 1000$ ppm of oxygen, humidity, and nitrogen by purging, sufficient for secondary containment to suppress and eliminate fire or hydrogen generation. Because the enclosure is inerted with argon, it also introduces a controlled asphyxiation hazard from a leak in the glovebox.

Scenario F (Glovebox III) shown in \cref{fig:6scenarios} (F) is a fully-equipped glovebox or enclosure with oxygen, humidity, and nitrogen scrubbers. The primary or secondary containment can be enclosed by a negative pressure to avoid any radioactive breach if it is involved. This also provides additional margin for overpressure \cite{d2023lithium}. There is no exposure to hazards even if the primary containment is breached. However, scenario F requires most of the safety features and thus has the highest complexity index, as shown in \cref{tab:p_jC}. 

\begin{table}[t]
\footnotesize
\centering
\begin{tabular}{l|l|c|cccccc}
\hline 
\multirow{2}{*}{i} & \multirow{2}{*}{Hazards}  & \multirow{2}{*}{$h_i$}   & \multicolumn{6}{c}{$p_i^{(\mathcal H)}$} \\
 &  &  & A   & B  & C   & D   & E   & F   \\ \hline
1   & H2 (bulk H2O)        &  5       & 0.5 & 0   & 0   & 0.5 & 0   & 0    \\ 
2   & H2 (moisture)        &  4       & 1   & 0.5 & 0   & 0.5 & 0   & 0    \\
3   & Li Fire (O2)         &  3       & 1   & 0.5 & 0.5 & 0.5 & 0   & 0    \\
4   & Li Fire (N2)         &  2       & 1   & 0.5 & 0.5 & 0.5 & 0   & 0    \\
5   & Li Smoke (fire)      &  3       & 1   & 0.5 & 0.5 & 0.5 & 0   & 0    \\
6   & Asphyxiation       &  1       & 0   & 0   & 0   & 0.5 & 0.5 & 0.5  \\ 
\hline 
    & Hazard index, $\mathcal H$         &          & 14.5  & 6    & 4    & 9    & 0.5    & 0.5     \\
\hline
\end{tabular}
\caption{Hazard index, $\mathcal H$, and qualitative level of exposure to hazards, $p_i^{(\mathcal H)}$ for each scenario (A to F). The value is estimated according to \cref{eq:piH}. Here, $p_i^{(\mathcal H)} = 0$ means low probability of exposure to hazards after failure of primary containment, $p_i^{(\mathcal H)} = 0.5$ means possible exposure to hazards but under some level of automatic control, and $p_i^{(\mathcal H)} = 1$ means full exposure or high probability of hazard to occur upon failure of the primary containment. The impact $h_i$ is taken from \cref{tab:hazards}.}
\label{tab:p_iH}
\end{table}
\begin{table}[h!]
\footnotesize
\centering
\begin{tabular}{l|l|c|cccccc}
\hline
\multirow{2}{*}{j} & \multirow{2}{*}{Safety feature}  & \multirow{2}{*}{$c_j$}   & \multicolumn{6}{c}{$p_j^{(\mathcal C)}$} \\
  &                                                    &    & A & B & C   & D   & E & F   \\ \hline
1 & Watertight 2nd containment                         & 2  & 0 & 1 & 0   & 0   & 0 & 0   \\
2 & Airtight 2nd containment                           & 3  & 0 & 0 & 1   & 1   & 1 & 1   \\
3 & Wet scrubber                                       & 5  & 0 & 1 & 0   & 0   & 0 & 0   \\
4 & Dry scrubber ($\mathrm{O_2}/\mathrm{N_2}$)         & 4  & 0 & 0 & 0   & 1   & 0 & 1   \\
5 & Ventilation \& exhaust                             & 3  & 1 & 1 & 1   & 1   & 1 & 1   \\
6 & Dehumidifier                                       & 2  & 0 & 0 & 1   & 0.5 & 0 & 1   \\
7 & $\mathrm{H_2}$ detector                            & 1  & 1 & 1 & 0.5 & 0.5 & 1 & 0.5 \\
8 & $\mathrm{H_2O}/\mathrm{O_2}/\mathrm{N_2}$ analyzer & 1  & 0 & 0 & 1   & 1   & 1 & 1   \\ \hline
    & Complexity index, $\mathcal C$         &          & 4  & 11    & 9.5    & 12.5    & 8    & 13.5     \\ \hline
\end{tabular}
\caption{Complexity index, $\mathcal C$, and qualitative level of feature requirement, $p_j^{(\mathcal C)}$ for each scenario (A to F). The value is estimated according to \cref{eq:pjC}, in which $p_j^{(\mathcal C)} = 0$ means the feature is not needed for the setup, hence zero probability of appearance in the system. $p_j^{(\mathcal C)} = 0.5$ means optional and $p_j^{(\mathcal C)} = 1$ means necessary and crucial for the confinement. The impact $c_j$ is taken from \cref{tab:complexity}.}
\label{tab:p_jC}
\end{table}

\begin{figure}
    \centering
    \includegraphics[width=\linewidth]{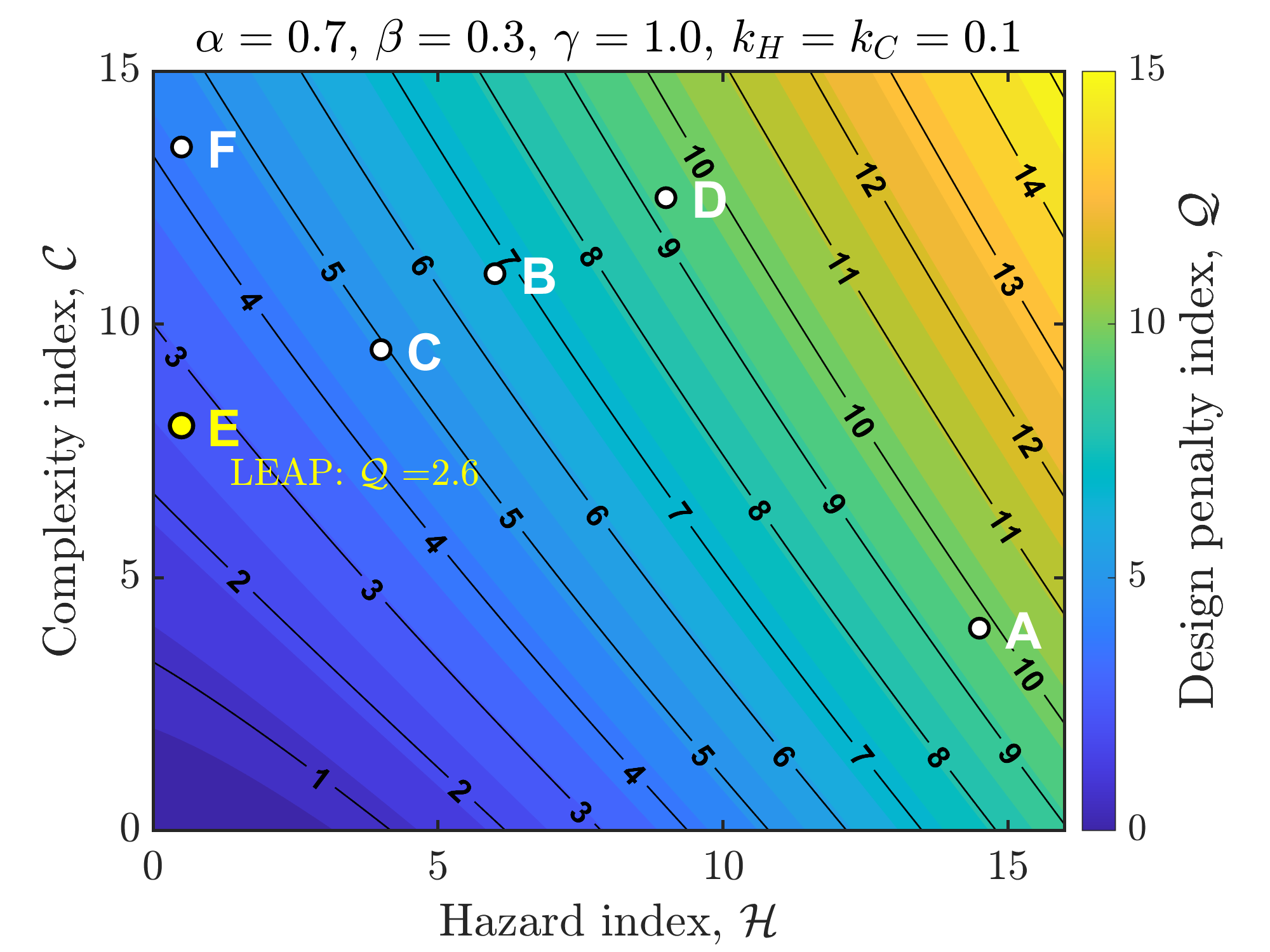}
    \caption{Design penalty index $\mathcal{Q}$ map for the six representative secondary containment scenarios. The contour shows $\mathcal{Q}=\alpha\mathcal{H}(1-\gamma_{\mathrm{eff}})+\beta\mathcal{C}$ with $\gamma_{\mathrm{eff}}=\gamma\exp(-k_H\mathcal{H}-k_C\mathcal{C})$, using $\alpha=0.7$, $\beta=0.3$, $\gamma=1.0$, and $k_H=k_C=0.1$. Points A--F correspond to the six containment scenarios based on existing liquid lithium systems listed in \cref{fig:6scenarios}. Scenario E, corresponding to the LEAP gloveroom architecture, gives the lowest penalty index, $\mathcal{Q}\approx2.6$, indicating a favorable balance between reduced lithium hazard exposure and moderate facility complexity. }
    \label{fig:Qcontour}
\end{figure}
%

\section{Lithium Experimental Application Platform (LEAP)}
\label{LEAP}

%
\begin{figure}[t]
    \centering
    \includegraphics[width=\linewidth]{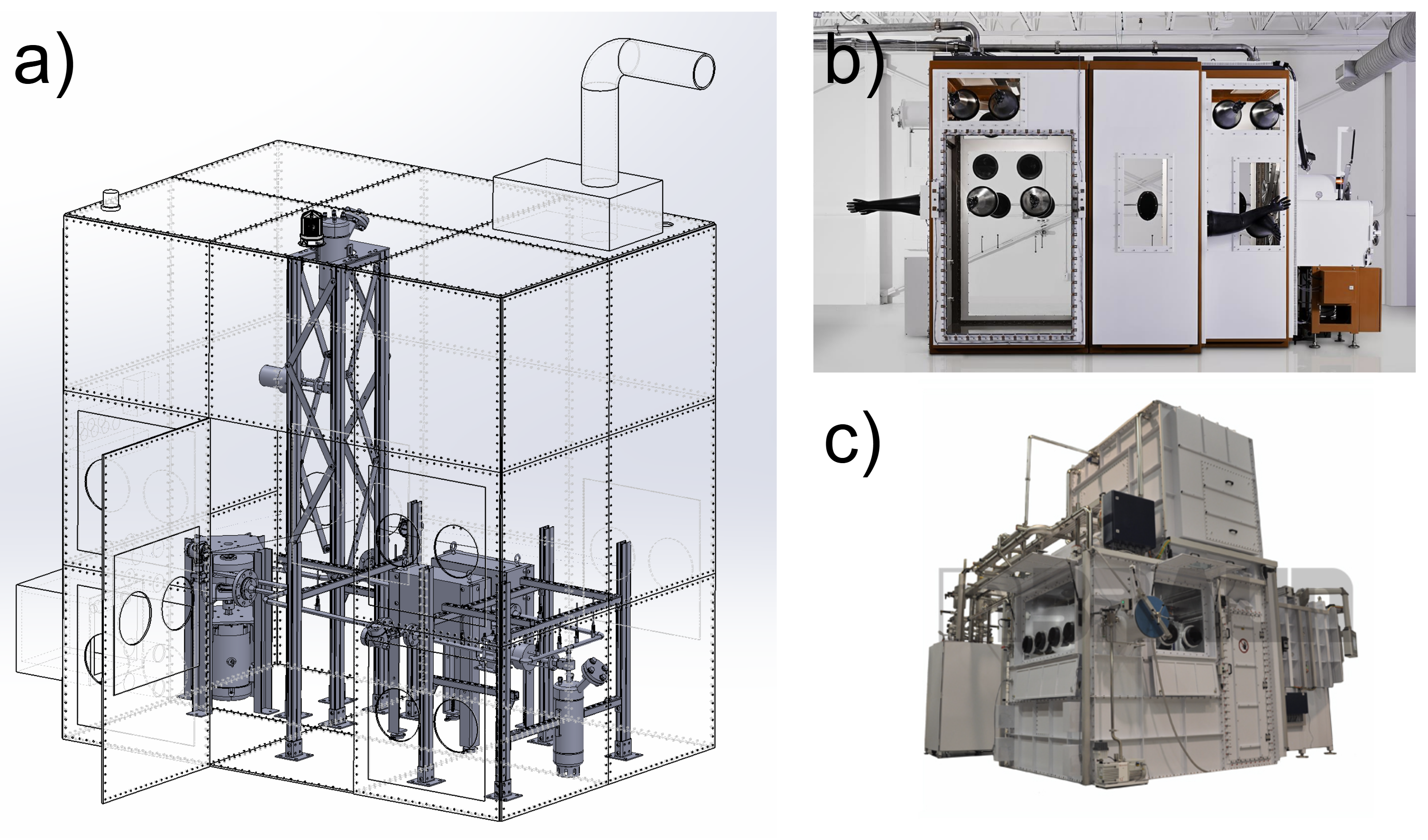}
    \caption{(a) A model of LEAP gloveroom with a Phase I liquid lithium loop inside. (b) Example product of a customized atmosphere-controlled enclosure. Courtesy of Inert Corp. (c) Another example product. Courtesy of M.Braun.}
    \label{fig:LEAPgloveroom}
\end{figure}

\subsection{Gloveroom containment solution}

We present the Lithium Experimental Application Platform (LEAP), currently under construction at Princeton Plasma Physics Laboratory, as a case study. LEAP is a modular, argon-inert liquid lithium test platform for flowing lithium PFC development, diagnostic validation, safety architecture evaluation, and future device integration studies. The initial system is designed for a lithium inventory of $\SI{22.68}{kg}$ ($\SI{50}{lb}$), remaining below the IBC's maximum allowable quantity of $\SI{45.36}{kg}$ ($\SI{100}{lb}$) for a control area using a ventilated enclosure.

\Cref{fig:Qcontour} summarizes the tradeoff between hazard reduction and implementation complexity for the six representative containment scenarios under the PPPL design preference used in this study. This example uses safety-prioritized weighting parameters, with $\alpha=0.7$, $\beta=0.3$, $\gamma=1.0$, and $k_H=k_C=0.1$. The contour field shows that the design penalty index $\mathcal{Q}$ increases with both hazard index $\mathcal{H}$ and complexity index $\mathcal{C}$, with a nonlinear contribution from the degradation of the effective mitigation coefficient. Scenario A is penalized by its high residual hazard despite its low complexity, whereas Scenarios B, C, D, and F reduce selected hazards by adding more infrastructure and operational controls. Scenario E, corresponding to the LEAP gloveroom architecture, lies in the low-hazard, moderate-complexity region and gives the lowest penalty value, $\mathcal{Q}\approx2.6$. Under the present design priorities, this supports the selection of an inert, airtight secondary enclosure around an enclosed lithium loop for research-scale flowing lithium systems.

The ranking in \Cref{fig:Qcontour} should be interpreted as a design-specific downselection result rather than a universal ordering of containment strategies. Different facilities may assign different values to $\alpha$, $\beta$, $\gamma$, $k_H$, and $k_C$ depending on lithium inventory, open-surface operations, tritium relevance, available infrastructure, staffing, and regulatory constraints. For example, a tritium processing system may justify the added complexity of Scenario F, whereas a small gram-scale lithium experiment may accept a simpler architecture. The value of the framework is that these priorities can be made explicit and varied systematically, allowing the preferred containment architecture to follow the mission requirements rather than being prescribed \textit{a priori}.

The defining feature of LEAP is a room-scale, prefabricated glovebox that serves as the secondary containment boundary ($\SI{3.93}{m}\ L \times \SI{2.37}{m}\ W \times \SI{3.35}{m}\ H$). As shown in \cref{fig:LEAPgloveroom} (a), it is a commercially fabricated, fully customized, walk-in argon enclosure that houses the lithium loop while maintaining a continuously purged and monitored inert atmosphere during operation, with target oxygen, nitrogen, and moisture levels below $\SI{1000}{ppm}$. \Cref{fig:LEAPgloveroom} (b) and (c) show two examples of the modularized gloverooms used for additive manufacturing, nuclear, or pharmaceutical applications. The enclosure is sized to support lithium experiments with a high-field magnet, with fully customized interface panels for gas, vacuum, electrical, cooling, and diagnostic feedthroughs, as well as future connections to external vacuum vessels. Active circulation inside the enclosure improves gas mixing, temperature uniformity, and analyzer response, while monitoring of oxygen, moisture, temperature, and hydrogen provides operational awareness during both inert operation and access conditions. The gloveroom also integrates facility-level functions that are difficult to achieve with a conventional glovebox, including door access with oxygen threshold interlocks, active HVAC ($\SI{6.59}{kW}$, or $\SI{22500}{BTU/hr}$), viewing ports, HEPA filtration, and anchor points and support structures for the liquid lithium and magnetic field systems. The enclosure and loop interfaces have completed PPPL design review processes, providing a traceable design basis for lithium operation in a national laboratory environment. In this sense, the LEAP gloveroom is not only a protective enclosure but also a standardized experimental infrastructure that combines inert secondary containment, modular utility access, and device-integration capability for iterative, flowing lithium PFC development.

A phased development strategy is used for the LEAP flowing lithium loop, with each stage increasing the level of physics integration and operational complexity. Phase I is an enclosed loop containing $\SI{2.27}{kg}$ ($\SI{5}{lb}$) of lithium, with a test section passing through a uniform magnetic field of up to $\SI{0.5}{T}$. This phase will characterize the pressure and flow performance of the moving magnet pump, validate thermometry and velocimetry diagnostics, and demonstrate lithium flow at up to $\SI{16}{GPM}$ ($\SI{1}{L/s}$) under strong magnetic field conditions relevant to flowing lithium PFC and hydrogenic fuel transport concepts \cite{ono2017liquid}.

Phase II extends LEAP to porous or open-surface lithium flow with applied heating and magnetic field exposure, using inventories up to $\SI{22.68}{kg}$ ($\SI{50}{lb}$). This phase will require modifications to the piping and support structure to accommodate PFC-relevant test articles. The primary goal is to characterize open-surface liquid lithium PFC performance under combined thermal loading and magnetic field conditions, including MHD pressure drop, heat transfer scaling, surface stability, and lithium-induced erosion or compatibility effects.

Phase III will develop interface strategies between LEAP and existing fusion devices, such as NSTX-U, to enable a continuous flowing lithium PFC loop coupled to a divertor surface. This stage is intended to move from standalone loop validation toward device-integrated operation, where lithium delivery, recovery, diagnostics, safety controls, and secondary containment must operate as a coordinated system.

\begin{table}[t]
\centering
\begin{tabular}{lcc}
\hline
Parameter & Symbol & LEAP values \\
\hline
Hazard coefficient & $\alpha$ & 0.7 \\
Complexity coefficient & $\beta$ & 0.3 \\
Hazard index & $\mathcal{H}$ & 0.5 \\
Complexity index & $\mathcal{C}$ & 8.0 \\
Nominal mitigation effectiveness & $\gamma$ & 1.0 \\
Hazard sensitivity of effectiveness & $k_H$ & 0.1 \\
Complexity sensitivity of effectiveness & $k_C$ & 0.1 \\
Effective mitigation coefficient & $\gamma_{\mathrm{eff}}$ & 0.4 \\
Design penalty index & $\mathcal{Q}$ & 2.6 \\
\hline
\end{tabular}
\caption{Representative parameter values assigned to the LEAP architecture in the design penalty framework. LEAP is treated as Scenario E in Tables~\ref{tab:p_iH} and~\ref{tab:p_jC}. The values are semi-quantitative estimates intended for comparative downselection rather than probabilistic risk assessment.}
\label{tab:LEAP_penalty_parameters}
\end{table}

\subsection{Lithium system design and major components}
\begin{figure}[t]
    \centering
    \includegraphics[width=0.75\linewidth]{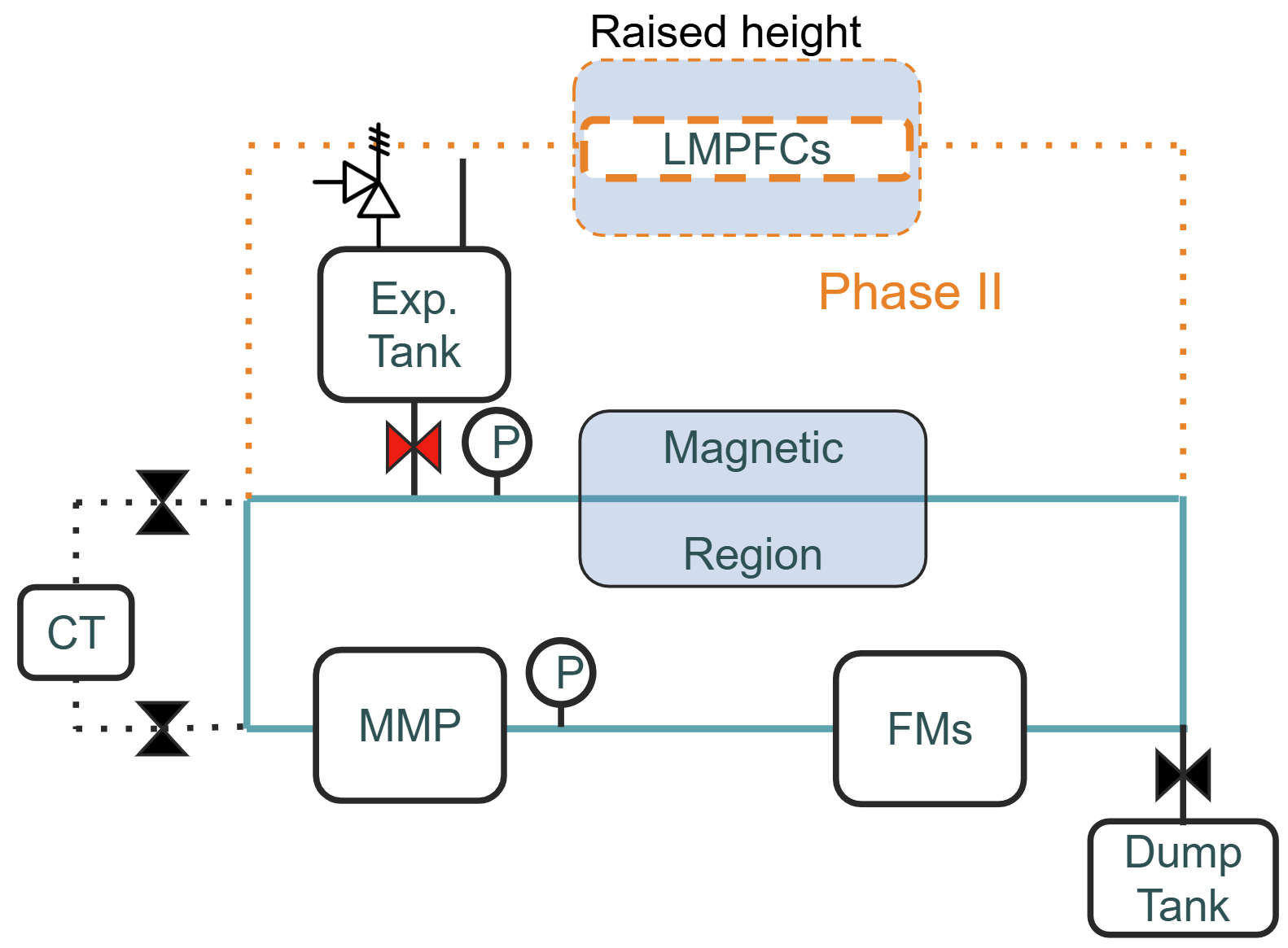}
    \caption{P\&ID of LEAP Phase I and II configuration, with potential expansion for purification by Cold Trap (CT). Phase I (in teal) is an enclosed lithium loop, while the expansion tank is the highest point in the system. Phase II (in orange) requires an open-surface flow region, which will be elevated to overcome cavitation and replace the expansion tank. }
    \label{fig:PID}
\end{figure}
\begin{figure}
    \centering
    \includegraphics[width=\linewidth]{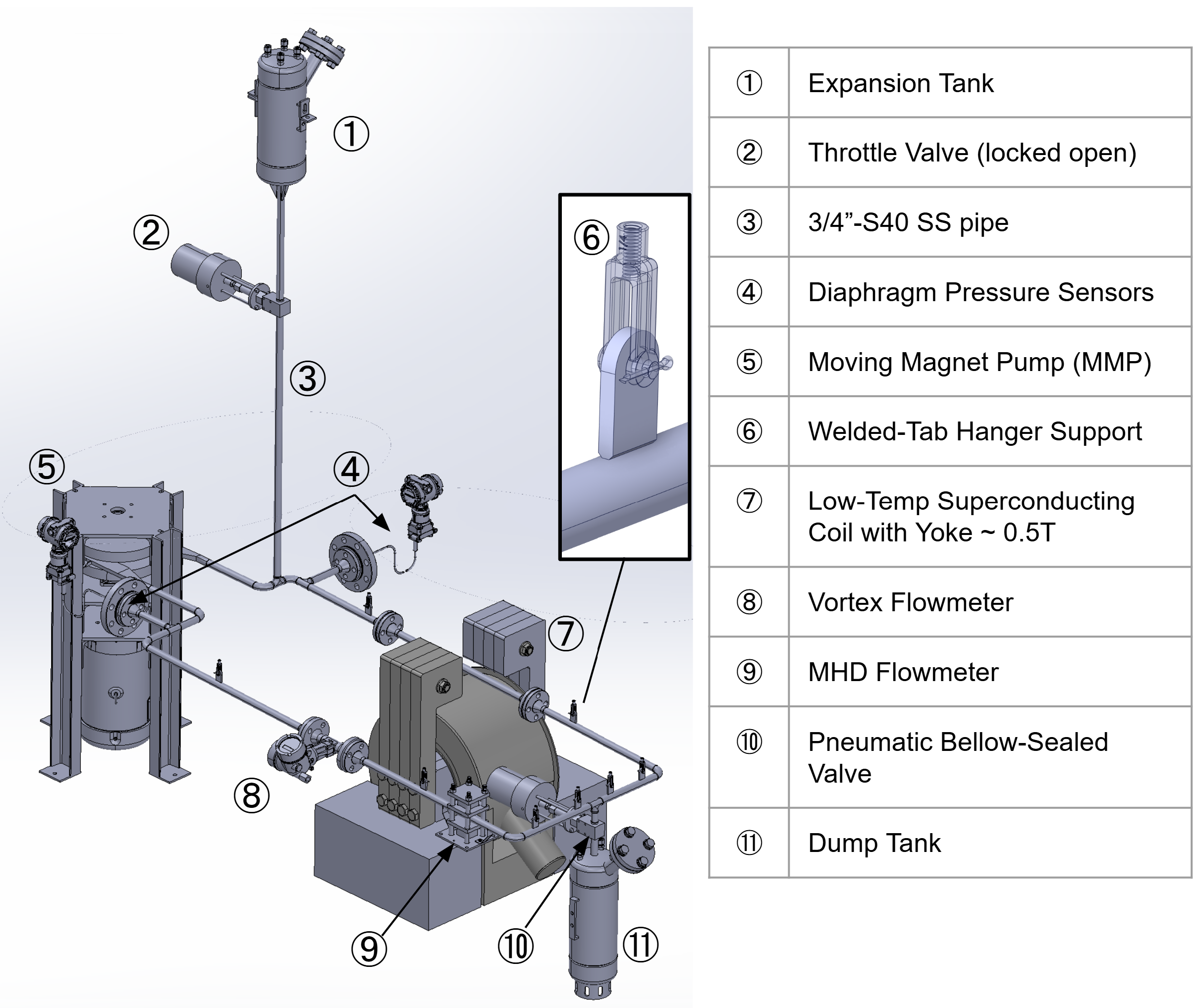}
    \caption{LEAP Phase I lithium loop and its major components.}
    \label{fig:phaseIloop}
\end{figure}

The LEAP lithium loop is designed as a modular flowing lithium system that can support enclosed pipe flow in Phase I and be modified for elevated porous or open-surface flow in Phase II. \Cref{fig:PID} shows the simplified process configuration. In Phase I, the loop is fully enclosed, with the expansion tank serving as the highest point with free-surface lithium in the system and the dump tank providing a low point for drainage and inventory recovery. In Phase II, the upper test region can be replaced or modified to support open-surface lithium PFC test articles at elevated height, while retaining the same general supply, drain, pumping, magnetic field, and diagnostic infrastructure. This staged configuration allows the platform to first validate lithium circulation, pump performance, diagnostics, pressure control, and MHD behavior before introducing more complex PFC-relevant geometries.

The major components of the Phase I loop are shown in \cref{fig:phaseIloop}. All lithium-wetted structural components are selected for lithium compatibility at elevated temperature, with a design operating envelope up to $\SI{400}{^\circ C}$. The primary flow boundary is constructed from $3/\SI{4}{in}$ Schedule 40 316 stainless steel piping, with the piping design guided by ASME B$31.3$ \cite{ASME-B31.3-2024}. The dump tank and expansion tank, designed in accordance with ASME BPVC \cite{ASME-BPVC-2025}, provide inventory management, fill-and-drain capability, pressure control, and accommodation for lithium thermal expansion. The expansion tank is placed at an elevated location to provide hydrostatic head and external pressure and to prepare the system for later Phase II open-surface operation, while the dump tank remains the primary collection volume for safe draining and cooldown. The loop has a slight tilt ($2^\circ$) towards the dump tank for natural drainage without power, as recommended in the sodium loop design \cite{jackson1955_liquid_metals_handbook_sodium_nak_supp}. 

Lithium circulation is driven by a moving magnet pump (MMP), which is designed to provide $\SI{16}{GPM}$ at a pressure rise of approximately $\SI{10}{psid}$ ($\SI{68.95}{kPa}$) by contactless pumping through the stainless steel flow channel. A locked-open throttle valve is included for flow control towards the expansion tank to protect expansion feedthroughs from abrupt overpressure, and pneumatic bellows-sealed stainless steel valves are used to isolate the main loop and dump tank. The test section passes through a cryogen-free low-temperature superconducting magnet with a low-carbon steel yoke capable of producing a transverse magnetic field of up to $\SI{0.5}{T}$, allowing the loop to investigate MHD pressure drop and flow behaviors with flexible space and under fusion-relevant magnetic field conditions.

The diagnostic package is designed to provide redundancy and cross-calibration on the measurements of lithium flow, temperature, and pressure. A vortex flowmeter and an MHD flowmeter are included for flow measurement. High-temperature diaphragm pressure gauges measure pressure differentials across the pump and flow path, while thermocouples or fiber optic temperature sensors provide distributed thermal monitoring of the loop, tanks, and heated sections. Together, these components allow LEAP to operate as both a flowing lithium test stand and a diagnostic validation platform for future liquid lithium PFC systems. Moreover, an additional purification loop can be added, such as the cold trap shown in \cref{fig:PID}.

Using the lithium properties at desired working condition at $\SI{200}{^\circ C}$ \cite{davison1968compilation}, including density $\rho=\SI{514.685}{kg/m^3}$, dynamic viscosity $\mu=6.27\times10^{-4}~\mathrm{Pa\cdot s}$, and electrical conductivity $\sigma=3.89\times10^{6}~\mathrm{S/m}$, the Phase I loop spans a strongly magnetized liquid-metal pipe-flow regime. To avoid ambiguity, the pipe inner diameter is used as the characteristic length for all dimensionless parameters. For the $3/\SI{4}{in.}$ Schedule 40 pipe, $D=\SI{20.93}{mm}$, a flow range of $0 - \SI{16}{GPM}$ corresponds to $U\approx0-\SI{2.93}{m/s}$. With $B=0-\SI{0.5}{T}$, the diameter-based Hartmann number is $Ha=BD\sqrt{\sigma/\mu}\approx0-820$, and the Reynolds number is $Re=UD/\nu\approx0-5.0\times10^4$. At $B=\SI{0.5}{T}$, the interaction parameter is $N=\sigma B^2D/(\rho U)=Ha^2/Re\approx13.5-\infty$ over the same flow range, decreasing with increasing flow speed. At the nominal $\SI{16}{GPM}$ and $\SI{0.5}{T}$ operating point, $Ha\approx820$, $Re\approx5.0\times10^4$, and $N\approx13.5$, indicating that Lorentz forces strongly affect flow behaviors.

\begin{figure}
    \centering
    \includegraphics[width=1\linewidth]{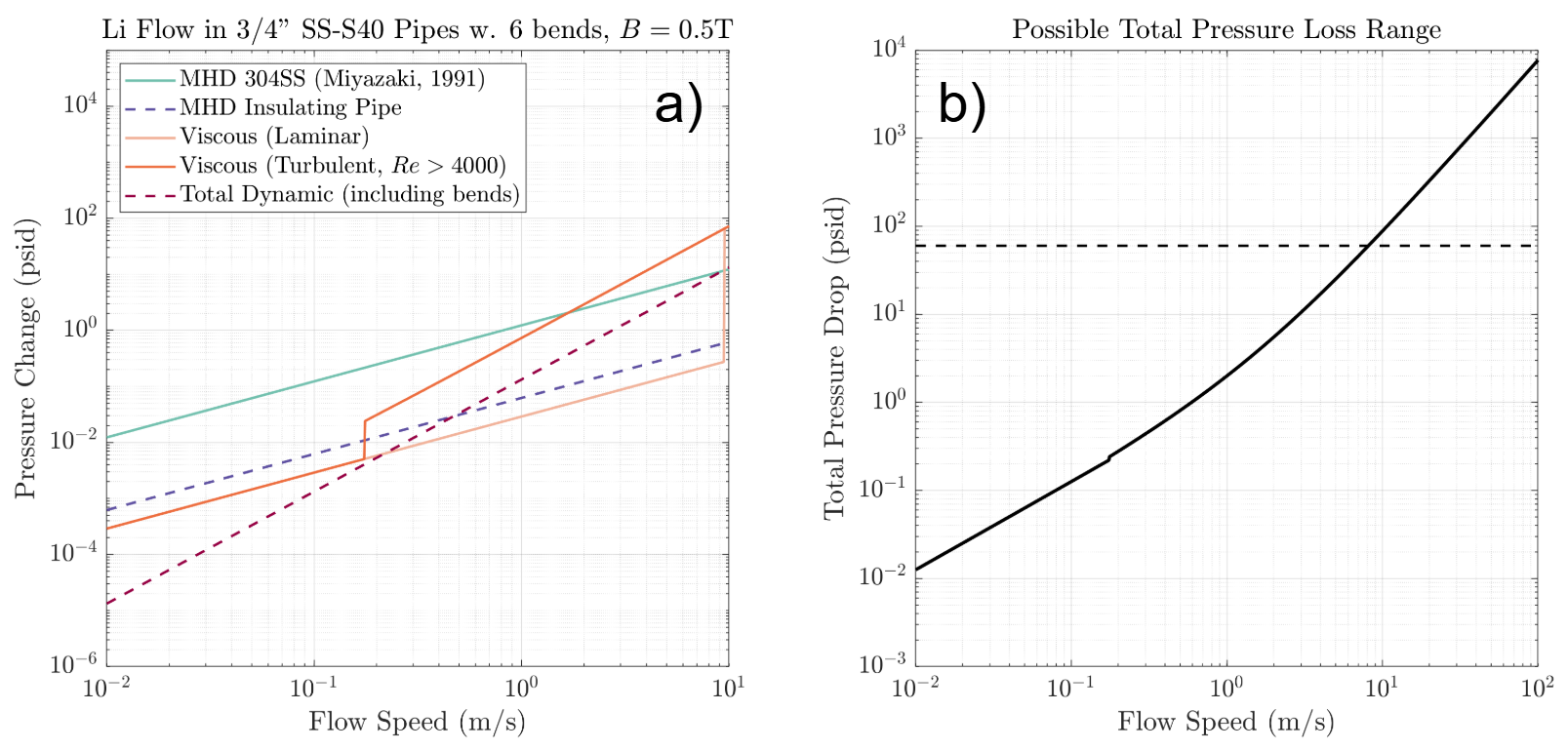}
    \caption{(a) Pressure drop due to MHD drag and viscous drag in various methods at $400^\circ$C, compared to the dynamic pressure across the entire pipe length of $\SI{6}{m}$. The length of pipe in $B=\SI{0.5}{T}$ magnetic field is $\ell = \SI{0.2}{m}$. Detailed derivation for each pressure is shown in \ref{app1}. (b) Total estimated pressure drop across the entire loop as a function of flow speeds. }
    \label{fig:pdrop}
\end{figure}

\Cref{fig:pdrop}(a) shows that MHD pressure drop is the dominant pressure sink in the Phase I loop over much of the operating range. Using the empirical estimate of Miyazaki \textit{et al.} ~\cite{miyazaki1991mhd} for liquid metal flow in a 304 stainless steel pipe, the MHD drag exceeds the laminar viscous pressure drop until the flow approaches the fully turbulent regime, expected at flow speeds between $\sim\SI{2}{m/s}$ and $\gtrsim\SI{10}{m/s}$ depending on the transition criterion. The total pressure drop, as shown in \cref{fig:pdrop}(b), remains below the $\SI{60}{psig}$ design operating pressure at the maximum intended flow speed and is compatible with the available pressure head of the MMP.

Additional comprehensive pressure, thermal, and structural analyses have been carried out during the design process, further validating the reliability of the LEAP lithium loop Phase I design. Selected analyses are shown in \ref{app2}, including the pressure balance and heat-loss estimates for the lithium loop.

\section{Discussion}
\label{discussion}

%
%
%

\begin{figure}
    \centering
    \includegraphics[width=1\linewidth]{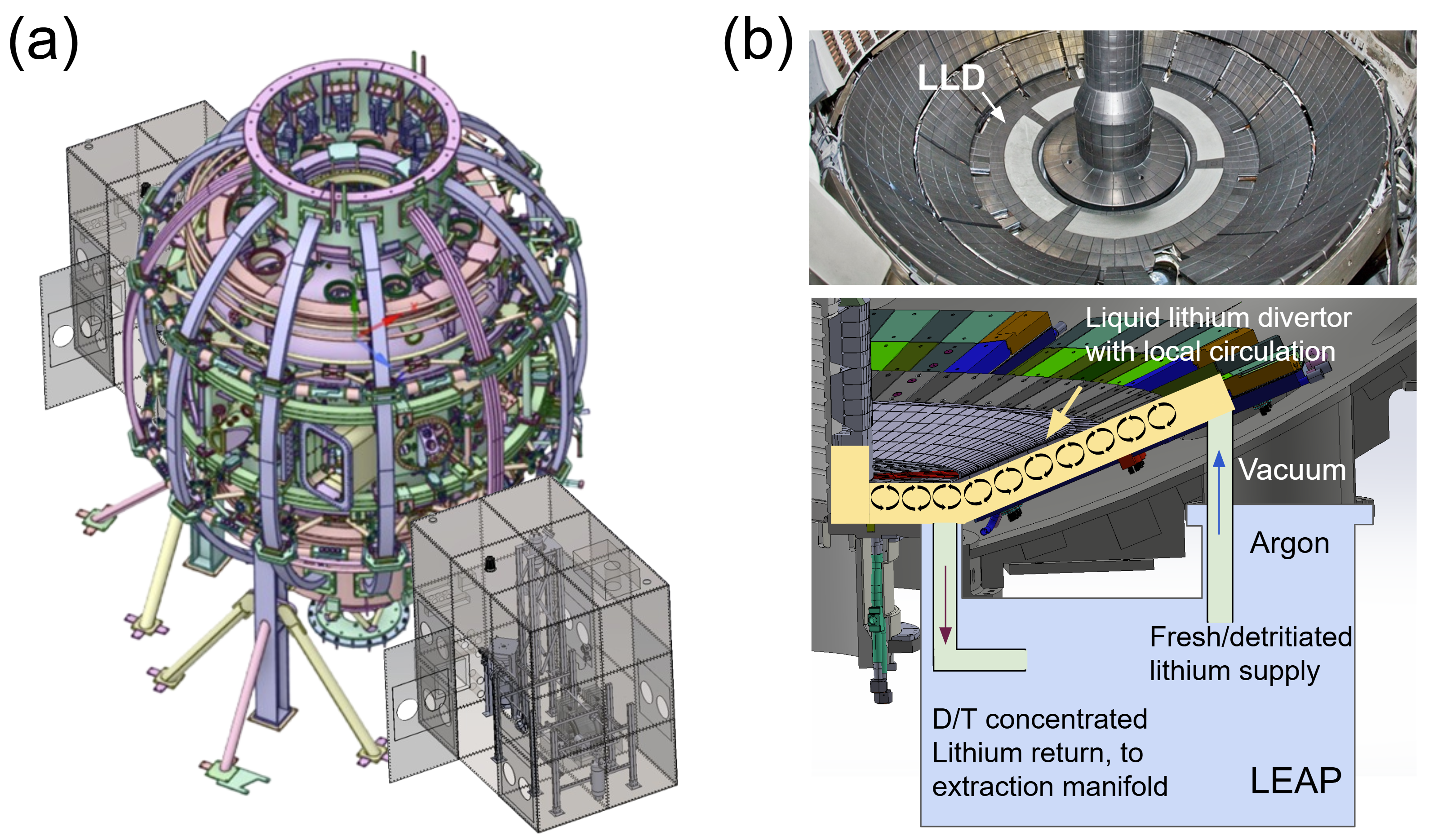}
    \caption{(a) Conceptual drawing with two LEAP gloverooms around the vacuum vessel in NSTX-U Test Cell. The exterior structure of the modular gloveroom can be modified to save space. Location of the gloverooms is arbitrarily chosen for illustration purposes only. (b) Conceptual LEAP interface with NSTX-U. LEAP enables a flexible interfacing by argon-to-vacuum feedthroughs with flange connections to the interior of NSTX-U's lower divertor region, where LLD segments were placed. Photo of LLD is adapted from H. Kugel \textit{et al.}, Fus. Eng. Design (2012) \cite{kugel2012nstx}. Liquid lithium divertor is coupled to a supply/drain system contained in LEAP's gloveroom system. Even though no tritium operation is planned for NSTX-U, the return loop can be used for pumping deuterium and tritium to the extraction manifold in Fusion Power Plants (FPPs) with a similar method. The cleaned lithium will re-enter the supply line. }
    \label{fig:NSTX_LEAP}
\end{figure}

\subsection{Integration of LEAP with fusion device}

The LEAP architecture by itself should be distinguished from plant-scale blanket loops and full fuel-cycle test facilities. LEAP is designed as a research-scale, modular secondary-containment platform for flowing lithium PFC development, diagnostic validation, MHD flow studies, safety architecture evaluation, and future device-interface concepts. It is therefore not intended to replace industrial-scale blanket or tritium fuel-cycle facilities, such as integrated lithium-lead and fuel-cycle test systems being developed for plant-relevant blanket and tritium-processing demonstrations. At larger plant scale, other containment strategies, including double-wall piping, leak-before-break design philosophy, guard vessels, segmented drain tanks, inerted pipe chases, and dedicated tritium confinement systems, may be more appropriate than a room-scale gloveroom. The value of LEAP is instead to provide a flexible intermediate platform and vacuum interface where lithium PFC concepts, flow control, diagnostics, hydrogen-isotope uptake and recovery concepts, and confinement-device interfaces can be developed before translation to larger blanket or fuel-cycle infrastructure.

The National Spherical Torus Experiment-Upgrade (NSTX-U) at the Princeton Plasma Physics Laboratory has features that are useful for a practical demonstration of integrating LEAP into a fusion device. The NSTX-U machine is equipped with $\SI{15}{MW}$ of neutral beam injection (NBI) and $\SI{6}{MW}$ of ion cyclotron resonance heating (ICRH), which can provide high heat loads to the divertor.

Because NSTX-U is a spherical tokamak, the surface area of the divertor is relatively small compared with those of large aspect ratio tokamaks with similar plasma heating capabilities. This practical advantage was demonstrated with the liquid lithium divertor (LLD) in NSTX. The NSTX device was the predecessor of NSTX-U and had the same divertor surface area.

The LLD was located on the outboard divertor as shown in \cref{fig:NSTX_LEAP}. It consisted of four segments with a radial width of $\SI{22}{cm}$, separated by a row of graphite tiles that included a radial array of Langmuir probes for measurements at the divertor strike points. There was also a toroidal gap that electrically isolated the inboard and outboard divertors. Each LLD segment consisted of a $\SI{0.25}{mm}$ stainless steel sheet that was dynamically bonded to a $\SI{2.2}{cm}$ thick copper substrate. A $\SI{0.17}{mm}$ layer of molybdenum was plasma-sprayed onto the stainless steel. This formed a surface with $45\%$ porosity for retaining static liquid lithium, which was kept liquefied with heaters in the copper substrate.

The LEAP system is intended to provide flowing liquid lithium from the outboard edge of the outboard divertor to the toroidal gap. Because the LEAP gloverooms are modular, their number around the NSTX-U vacuum vessel can be adjusted to meet lithium inventory, pumping power, and flow required by any flowing liquid lithium divertor design. A possible configuration for the LEAP gloverooms in the NSTX-U Test Cell and a conceptual interface with the vacuum vessel are shown in \cref{fig:NSTX_LEAP}.

The modularity of the LEAP system is itself a safety feature. Each LEAP gloveroom and the NSTX-U vacuum vessel is expected to accommodate only their individual lithium inventories. In that way, no component has to meet the safe handling requirements for the entire lithium inventory. 

Each LEAP gloveroom also feeds liquid lithium into a common manifold inside NSTX-U vacuum vessel. This redundancy allows NSTX-U to be provided with liquid lithium if any LEAP gloveroom becomes inoperable.

\subsection{Limitations and considerations of the LEAP lithium loop and gloveroom}

LEAP's Phase II lithium loop includes an open-surface flow for testing plasma-facing components. One of the major challenges for many open-surface lithium flow systems is avoiding the cavitation effect due to lithium's relatively high vapor pressure relative to other metals. Cavitation happens when local static pressure ($P_0-P_d$) is smaller than the vapor pressure of the liquid,
\begin{equation}
    P_{H}-\frac{1}{2}\rho U^2 \lesssim P_V.
\end{equation}
The cavitation number, $Ca$, is the ratio of a flowing liquid's local static and dynamic pressure.
\begin{equation}
    Ca = \frac{P_H-P_V}{\frac{1}{2}\rho U^2},
\end{equation}
where in an open-surface liquid lithium system, $P_H = \rho g \Delta h$, $\Delta h$ is the relative system height of the Phase II loop, and $P_d$ is the dynamic pressure. To avoid the cavitation effect, $Ca \gg 1$ should be maintained at the pump head. Therefore, a sufficient hydrostatic head is preferred for open-surface flow. At $400^\circ$C, lithium has a vapor pressure of $0.011~$Pa \citep{davison1968compilation}. 
\begin{equation}
    \log(P_V) = 10.015 - \frac{8064.5}{T} = 10.015 - \frac{8064.5}{673.15},\quad P_V = 0.011\,  \text{Pa},
\end{equation}
where $T$ is the lithium temperature in Kelvin. 
\begin{figure}[t]
    \centering
    \includegraphics[width=\linewidth]{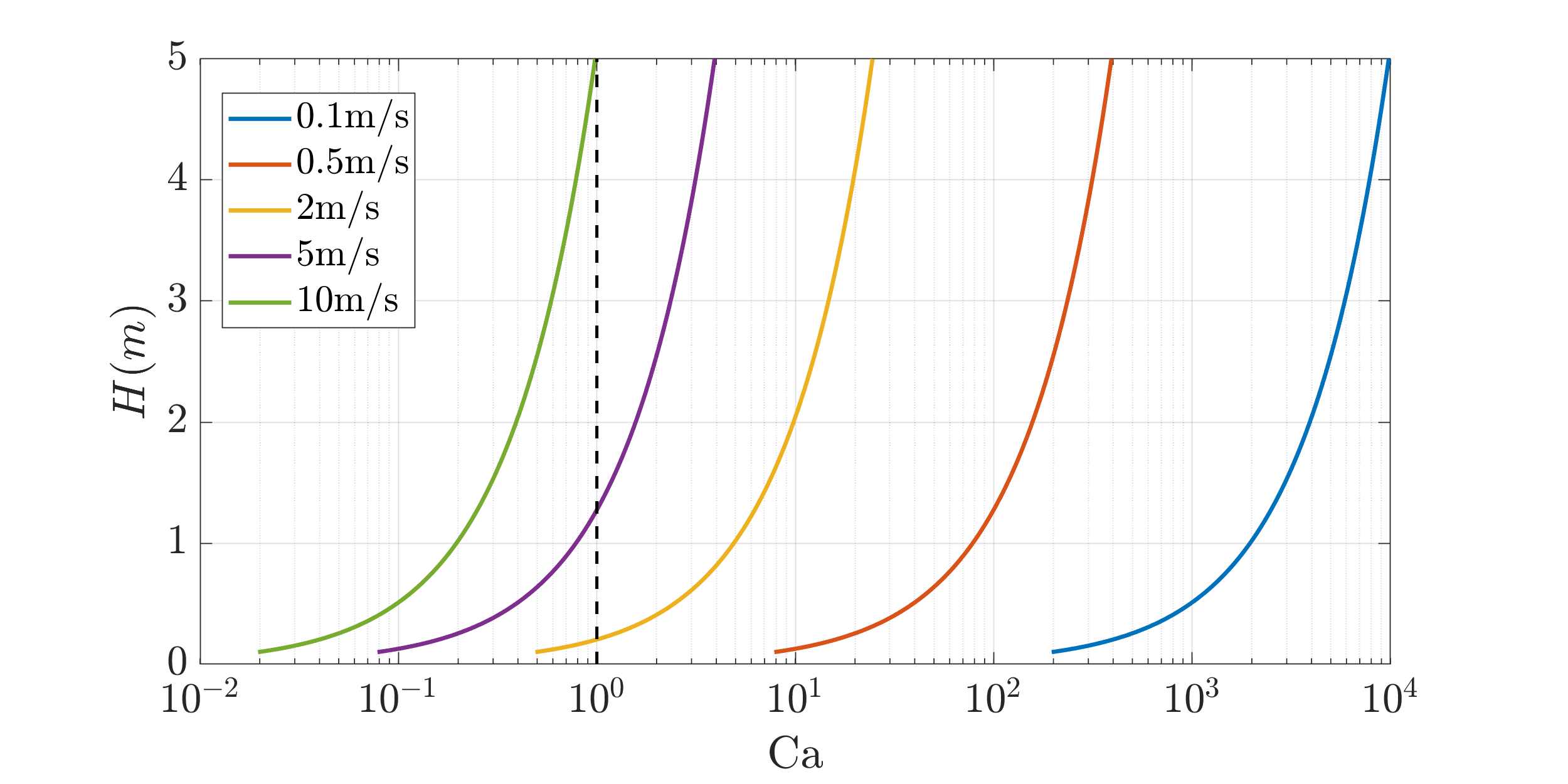}
    \caption{System heights vs. cavitation number at the pump inlet, showing that faster lithium flow requires a higher system height to prevent cavitation effects at the pump inlet. A $2$m-tall system running $\SI{2}{m/s}$ flow has a $Ca \approx 10$, which can be considered as cavitation safe. We consider $\SI{5}{m/s}$ flow reaches $Ca = 2$ at a system height of $\SI{2.55}{m}$ as the design limit.}
    \label{fig:cav}
\end{figure}

\Cref{fig:cav} shows cavitation number as a function of system height differences with different flow speeds. A fast flow system with several meters per second flow must have a considerable system height in the order of meters. For this reason, the current design of LEAP's gloveroom has a height of $3.35~$m.  The cavitation effect limits the maximum flow speed and provides a guideline for determining system height of the Phase II system. For instance, $Ca = 10$ is achieved for a $2~$m-tall system flowing $2~$m/s, which could be a desired operating parameter for the LEAP Phase II system. 

However, increasing the system height is not always practical. To keep the secondary containment compact and limit argon consumption during purging, the pipe diameter before the pump inlet can be locally increased. A larger pipe reduces the flow velocity and increases the cavitation number, but it also increases the total lithium inventory.

The LEAP gloveroom provides a practical secondary containment solution for flowing lithium experiments, but its room-scale argon enclosure also introduces operational limitations. The most important is purge time and argon consumption. For a conventional sweep-through purging, the required inert-gas volume can be estimated from the complete-mixing relation $\dot{V}t=V\ln(C_0/C_t)$, with a correction factor $K$ used to account for nonideal mixing \cite{kinsley2001properly}. $K = 1$ is perfect mixing. The argon purge time is 
\begin{equation}
    t = \frac{V}{K\dot{V}}\ln\left(\frac{C_0}{C_t}\right),
\end{equation}
where $V$ is the enclosure volume, $\dot{V}$ is the argon purge rate, $C_0$ is the initial oxygen concentration, and $C_t$ is the target concentration. For $V\approx\SI{12}{m^3}$ ($\SI{424}{ft^3}$), $\dot{V}=\SI{10}{scfm}$, $C_0=\SI{210000}{ppm}$, and $C_t=\SI{1000}{ppm}$, the ideal complete-mixing estimate gives $t\approx\SI{3.8}{h}$ and $\sim\SI{2300}{scf}$ of argon. With a mixing correction factor of $K=0.5$, the estimate increases to $t\approx\SI{7.6}{h}$ and $\sim\SI{4500}{scf}$ of argon. A standard 230-liter argon dewar (liquid cylinder) typically provides approximately $6000$ standard cubic feet (SCF). The gloveroom targets $\SI{1000}{ppm}$ oxygen, nitrogen, and moisture, rather than glovebox grade impurity levels near $\sim\SI{1}{ppm}$, as the gloveroom is intended only to provide inert secondary containment, not ultra-high purity materials processing. 

This constraint can be alleviated by the slow argon displacement technique, with gas inlets placed near the floor so denser argon can slowly and effectively displace air upward toward the exhaust path, reducing total argon usage to several times the volume of gas inside the gloveroom. After the target atmosphere is reached, active PID-controlled circulation can then be initiated and used to homogenize gas composition, improve temperature uniformity, and provide representative analyzer responses. A liquid argon dewar is preferred for extended campaigns because it reduces cylinder changeout burden and better supports slow purge, cover gas operation, and repeated inerting cycles. Moreover, flexible enclosure geometry conforming to the contour of the lithium system using modularized enclosures and compartments can effectively reduce the enclosed volume.

The second limitation is access. Since personnel entry is incompatible with an oxygen-deficient argon atmosphere, maintenance must be planned around purges. The third limitation is scale: increasing the enclosure size would improve access and support taller Phase II open-surface flow geometries, but would also increase purge volume, argon demand, HVAC load, and atmosphere control burden. These limitations motivate the staged LEAP approach, in which Phase I validates enclosed pipe flow, gas handling, monitoring, and thermal management before larger inventory, open surface, and device interface operations are introduced.

Beyond fusion, liquid-metal systems are used in metallurgy \cite{wei2008centrifugal}, pharmaceuticals \cite{wang2018preparations}, soft and flexible electronics \cite{dickey2017stretchable}, and laboratory studies of astrophysical and geophysical magnetohydrodynamics \cite{adams2015liquid,stefani2019dresdyn,xu2022thermoelectric,liu2026geostrophic}. Despite differences in working fluid, temperature, and hazard profile, these systems share a common challenge: reactive, conductive, or high-temperature fluids must be contained without imposing impractical complexity. The containment logic and design penalty index developed here can therefore be adapted to other liquid-metal systems, particularly alkali-metal systems, where chemical reactivity, inerting, diagnostics, and operational access must be balanced.

\section{Summary and conclusions}
\label{conclusion}

This work presents a risk-informed design framework for selecting secondary containment architectures for flowing liquid lithium systems. Rather than treating safety and facility complexity as independent objectives, the framework evaluates their qualitative tradeoff through a design penalty index. When applied to representative containment scenarios under PPPL design requirements, the analysis identifies an argon-purged enclosure around an enclosed lithium loop as a favorable architecture for lithium PFC development. 

The LEAP case study demonstrates how this tradeoff can be implemented in practice. A modular argon gloveroom provides an inert secondary containment boundary that suppresses the dominant lithium reaction pathways while preserving access, diagnostics, hardware flexibility, and future device interface capability. The architecture also introduces practical constraints, including purge time, argon consumption, oxygen-deficiency controls, and access planning. These costs are acceptable for a staged research platform that must progress from enclosed pipe flow to open-surface lithium PFC testing and eventual fusion-device integration.

Containment architecture should be treated as part of the technology development pathway, not merely as supporting infrastructure. For flowing lithium PFC test systems, a modular inert gloveroom offers a practical middle ground between bench-scale gloveboxes and dedicated lithium facilities. More broadly, the hazard/complexity framework provides a general method for adapting lithium system designs to different inventories, operating modes, facility constraints, and fusion applications. The same logic can also inform other liquid metal systems where safety, maintainability, diagnostics, and rapid iteration must be balanced.

\section*{Acknowledgements}
The research described in this paper was conducted under the Laboratory Directed Research and Development (LDRD) Program at Princeton Plasma Physics Laboratory, a national laboratory operated by Princeton University for the U.S. Department of Energy under Prime Contract No. DE-AC02-09CH11466. The United States Government retains a non-exclusive, paid-up, irrevocable, worldwide license to publish or reproduce the published form of this manuscript, or allow others to do so, for United States Government purposes.

The authors thank Richard Majeski, Robert Ellis, Dang Cai, Andrei Khodak, Christian Veyssiere, Andrew Shone, Elise Demoncheaux, Enrique Miralles-Dolz, the PPPL ESH team, and the LEAP design review committee for fruitful discussions and feedback. The authors thank Rich Cavanaugh, Jiawen Wang, and the rest of the engineering team for LEAP-related design, drawing, and analysis work. The authors thank Creative Engineers, Inc. for comprehensive lithium safety training, which led to many considerations in this work. 

During the preparation of this work, the authors acknowledge the use of Writefull and ChatGPT (OpenAI) for language editing and improving readability. After using this tool/service, the authors reviewed and edited the content as needed and take full responsibility for the content of the publication.

\appendix

\section{Optimization of the design penalty index}
\label{app1}
For convenience, the design penalty indices for different design options are written as
\begin{equation}
    \mathcal{Q}(\mathbf{x})
    =
    \alpha \mathcal{H}\left(1-\gamma_{\mathrm{eff}}(\mathbf{x})\right)
    +
    \beta \mathcal{C}(\mathbf{x}),
    \label{eq:Qopt0}
\end{equation}
with
\begin{equation}
    \gamma_{\mathrm{eff}}(\mathbf{x})
    =
    \gamma(\mathbf{x})\exp\!\left(-k_H \mathcal{H}(\mathbf{x})-k_C \mathcal{C}(\mathbf{x})\right),
    \label{eq:gammaeffopt0}
\end{equation}
and
\begin{equation}
    \gamma(\mathbf{x})
    =
    1-\exp\!\left(-\sum_{m}\eta_m x_m\right).
    \label{eq:gammaopt0}
\end{equation}

The optimization problem is
\begin{equation}
    \min_{\mathbf{x}\in {X}} \mathcal{Q}(\mathbf{x}),
    \label{eq:opt0}
\end{equation}
where $X$ denotes the admissible design set.

If the design features are binary (absent/present), \(x_m\in\{0,1\}\), which is suggested in \cref{method}, the problem is not solved by continuous first-order conditions. Instead, the actual optimum is obtained by discrete search, as used in our current study. This search is practical when the number of candidate safety features is modest. However, for the completeness of the problem, we hereby include continuous design features (e.g., argon purge rate, oxygen and moisture set point, spill-capture capacity, redundancy level) and demonstrate a continuous relaxation, which can be useful for ranking feature sensitivity. 

\subsection{General first-order condition}

Let
\begin{equation}
    E(\mathbf{x})=\exp\!\left(-k_H \mathcal{H}(\mathbf{x})-k_C \mathcal{C}(\mathbf{x})\right),
\end{equation}
so that
\begin{equation}
    \gamma_{\mathrm{eff}}(\mathbf{x})=\gamma(\mathbf{x})E(\mathbf{x}).
\end{equation}
Then the design penalty indices become
\begin{equation}
    \mathcal{Q}(\mathbf{x})
    =
    \alpha \mathcal{H}(\mathbf{x})\left(1-\gamma(\mathbf{x})E(\mathbf{x})\right)
    +
    \beta \mathcal{C}(\mathbf{x}).
\end{equation}

Differentiating with respect to a generic and continuous design variable \(x_m\) gives
\begin{align}
    \frac{\partial \mathcal{Q}}{\partial x_m}
    &=
    \alpha \frac{\partial \mathcal{H}}{\partial x_m}\left(1-\gamma E\right)
    -
    \alpha \mathcal{H}\frac{\partial (\gamma E)}{\partial x_m}
    +
    \beta \frac{\partial \mathcal{C}}{\partial x_m}.
    \label{eq:dQdx_general_1}
\end{align}
Applying the product rule,
\begin{equation}
    \frac{\partial (\gamma E)}{\partial x_m}
    =
    E\frac{\partial \gamma}{\partial x_m}
    +
    \gamma \frac{\partial E}{\partial x_m},
\end{equation}
and
\begin{equation}
    \frac{\partial E}{\partial x_m}
    =
    -E\left(
    k_H\frac{\partial \mathcal{H}}{\partial x_m}
    +
    k_C\frac{\partial \mathcal{C}}{\partial x_m}
    \right).
\end{equation}
Therefore,
\begin{equation}
    \frac{\partial (\gamma E)}{\partial x_m}
    =
    E\frac{\partial \gamma}{\partial x_m}
    -
    \gamma E\left(
    k_H\frac{\partial \mathcal{H}}{\partial x_m}
    +
    k_C\frac{\partial \mathcal{C}}{\partial x_m}
    \right).
\end{equation}
Substituting into \cref{eq:dQdx_general_1} yields
\begin{equation} \label{eq:dQdx_general_final}
    \begin{split}
    \frac{\partial \mathcal{Q}}{\partial x_m}
    &=
    \alpha \frac{\partial \mathcal{H}}{\partial x_m}\left(1-\gamma E\right)
    -\alpha \mathcal{H}E\frac{\partial \gamma}{\partial x_m}\\
    &+\alpha \mathcal{H}\gamma E
    \left(
    k_H\frac{\partial \mathcal{H}}{\partial x_m}
    +
    k_C\frac{\partial \mathcal{C}}{\partial x_m}
    \right)
    +\beta \frac{\partial \mathcal{C}}{\partial x_m}.
    \end{split}
\end{equation}
From \cref{eq:gammaopt0},
\begin{equation}
    \frac{\partial \gamma}{\partial x_m}
    =
    \eta_m \exp\!\left(-\sum_n \eta_n x_n\right)
    =
    \eta_m(1-\gamma).
    \label{eq:dgammadx}
\end{equation}
Substituting \cref{eq:dgammadx} into \cref{eq:dQdx_general_final} gives
\begin{equation}
\begin{split}
    \frac{\partial \mathcal{Q}}{\partial x_m}
    &=
    \alpha \frac{\partial \mathcal{H}}{\partial x_m}\left(1-\gamma E\right)
    -\alpha \mathcal{H}E\,\eta_m(1-\gamma) \\
    &+ 
    \alpha \mathcal{H}\gamma E
    \left(
    k_H\frac{\partial \mathcal{H}}{\partial x_m} 
    +
    k_C\frac{\partial \mathcal{C}}{\partial x_m}
    \right)
    +\beta \frac{\partial \mathcal{C}}{\partial x_m}.
    \label{eq:dQdx_with_gamma}
\end{split} 
\end{equation}

An interior optimum \(\mathbf{x}^\ast\) must satisfy
\begin{equation}
    \frac{\partial \mathcal{Q}}{\partial x_m}\Big|_{\mathbf{x}^\ast}=0
    \qquad
    \text{for all }m.
    \label{eq:foc_general}
\end{equation}
\subsection{Special case: fixed \texorpdfstring{$\mathcal{H}$}{H}, linear \texorpdfstring{$\mathcal{C}$}{C}}

A practically useful simplification is to treat the intrinsic hazard level as fixed for a given experiment class,
\begin{equation}
    \mathcal{H}(\mathbf{x})=\mathcal{H}_0,
    \qquad
    \frac{\partial \mathcal{H}}{\partial x_m}=0,
\end{equation}
and to model the complexity as a weighted linear sum,
\begin{equation}
    \mathcal{C}(\mathbf{x})=\sum_m c_m x_m,
    \qquad
    \frac{\partial \mathcal{C}}{\partial x_m}=c_m.
\end{equation}
Then \cref{eq:dQdx_with_gamma} reduces to
\begin{equation}
    \frac{\partial \mathcal{Q}}{\partial x_m}
    =
    -\alpha \mathcal{H}_0 E\,\eta_m(1-\gamma)
    +\alpha \mathcal{H}_0\gamma E\,k_C c_m
    +\beta c_m,
    \label{eq:dQdx_special}
\end{equation}
where
\begin{equation}
    E=\exp\!\left(-k_H\mathcal{H}_0-k_C\mathcal{C}(\mathbf{x})\right).
\end{equation}
This derivative indicates whether increasing the extent or strength of the mitigation feature $m$ would reduce the penalty in the continuous relaxation of the problem. Thus, the first-order condition becomes
\begin{equation}
    -\alpha \mathcal{H}_0 E\,\eta_m(1-\gamma)
    +\alpha \mathcal{H}_0\gamma E\,k_C c_m
    +\beta c_m
    =0.
    \label{eq:foc_special}
\end{equation}
The first term represents the hazard-reduction benefit of adding the feature $m$ and is largest when the intrinsic hazard is high, and the current mitigation level is still limited. The second term is complexity-induced degradation of effectiveness. The third term is the direct complexity penalty. This last penalty is independent of hazard reduction physics. It simply reflects that complex systems are harder and more expensive to build and run.

\Cref{eq:foc_special} has a clear interpretation: a safety feature should be included only up to the point where its reduction of residual hazard is matched by the added penalty from increased facility complexity and from the loss of practical effectiveness caused by that extra complexity.

\section{Pressure and Thermal Consideration for Phase I loop}
\label{app2}
\begin{figure}
    \centering
    \includegraphics[width=0.7\linewidth]{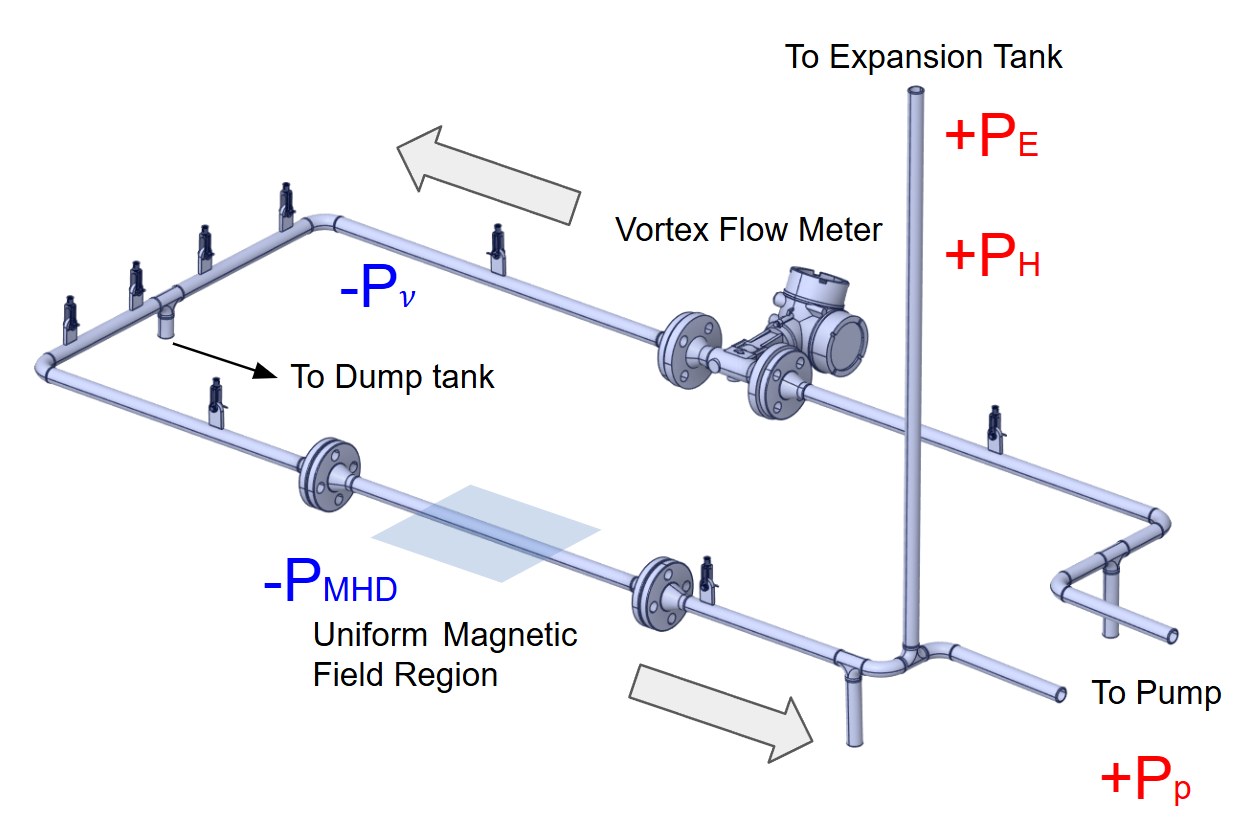}
    \caption{Schematics of the pressure balance for Phase I configuration, which has a main loop in the horizontal plane.}
    \label{fig:press}
\end{figure}

\Cref{fig:press} shows the pressure balance within the Phase I loop. The valve connected to the dump tank is closed during operation, and the flow-restriction throttle valve is locked on, so only the internal pressure of the main loop should be considered. The main loop, the moving magnet pump, denoted as `P' in the diagram, and the magnet are placed in the same horizontal plane. 
    
The main loop during operation is filled with liquid lithium, with the only open surface at the top of the expansion tank. Hydrostatic pressure $P_H = \rho g \Delta h$ and external pressure $p_E$ from the top of the expansion tank add a background pressure into the system, whereas pressure changes due to MHD drag $\Delta P_{M\!H\!D}$ from the piping region inside the strong magnetic field and viscous drag $\Delta P_{\nu}$ reduce the pressure from the pump $P_p$. Here, $g$ is the gravitational acceleration, and $\Delta h$ is the distance between the top fluid layer and the main loop level. The static fluid pressure at the pump inlet $P_{static}$ is likely the lowest pressure point across the entire main loop and has to be larger than zero (or the vapor pressure, see section 1.3) to drive a flow,
\begin{equation}
    P_{static} = P_H + P_E = P_E + P_p - (\Delta P_{M\!H\!D} + \Delta P_\nu + P_d + P_{diag})>0,
    \label{eq:I}
\end{equation}
where $P_E \approx P_{atm}$ is the external pressure close to the gloveroom pressure. \Cref{fig:pdrop} a) summarizes the value of the pressure drop terms for MHD drag $\Delta P_{M\!H\!D}$, viscous drag $\Delta P_\nu$, dynamic pressure $P_d$, and pressure drop across the diagnostics $P_{diag}$, as a function of flow speed. \Cref{fig:pdrop} b) shows the estimated total pressure drop by summing previous predictions of pressure-dropping terms. The designed operating pressure for the system is $\SI{60}{psig}$, marked in black horizontal dashed line. 

Clearly, the MHD pressure drop dominates other factors until the flow speed exceeds $\sim \SI{1.7}{m/s}$, at which point turbulent viscous drag could become more significant. At $\SI{60}{psig}$, the internal pressure can sustain a maximum of $\sim \SI{8}{m/s}$ flow. Moreover, there is no significant difference between $200^\circ$C and $400^\circ$C. 

\subsection{MHD drag}

There are several methods to calculate the MHD pressure drop for a circular conducting pipe flow perpendicular to a uniform magnetic field. The pressure drop from the MHD drag can be calculated via integrating of the Hartmann pressure gradient with an insulating circular boundary \citep{shercliff1962magnetohydrodynamic}. The wall conductance ratio $C_w = \sigma_w t_w/(\sigma d)$ indicates the conductivity ratio of the pipe to fluid. Here $t_w$ is the wall thickness. For $3/4$"- S40 stainless steel pipe $\lambda \approx \SI{2.87}{mm}$. At $200^\circ$C, the electrical conductivity of lithium is $\sigma \approx 3.89 \times 10^6~\SI{}{S/m}$, and $\sigma_w \approx 1.4\times 10^6~\SI{}{S/m}$ for stainless steel. Substituting lithium and stainless steel parameters, $C_w = 0.0175$ for our SS-lithium pipe flows, the stainless steel conducting wall does not significantly reduce MHD drag and can thus be treated as an insulating pipe. The pressure drop can be rewritten from eq.~9 of \cite{shercliff1962magnetohydrodynamic}, 
\begin{equation}
    \Delta P_{M\!H\!D} = \left( \frac{3\pi}{4}\right) \left(\frac{UB \ell \sqrt{\sigma \mu}}{D}\right) \left({\frac{2Ha}{2Ha-3\pi}}\right),
    \label{eq:MHD_pipe}
\end{equation}
where $U$ is the velocity of the fluid and $\sigma$ is the electrical conductivity. A strong magnetic field $B = \SI{0.5}{T}$ is present and perpendicular to the flow direction for a total length of $\ell = \SI{0.2}{m}$. At high $Ha$, $2Ha/(2Ha-3\pi) \rightarrow 1$. This MHD pressure drop is marked as the purple line in \cref{fig:pdrop} a). Alternatively, the simplest pressure drop for a high-Hartmann number flow with insulating boundaries is also given as \citep{davidson2017introduction}
\begin{equation}
    \Delta P_{M\!H\!D} \sim \frac{\sigma U B^2 \ell}{Ha},
    \label{eq:MHD_ins}
\end{equation}
which is marked as the blue line in \cref{fig:pdrop} a). It approximates the previous solution. Here, we use empirical results from MHD pressure change in a 304SS pipe \cite{miyazaki1991mhd} as an approximation to our 316SS pipes, 

\begin{equation}
    \Delta P_{M\!H\!D} = \left( \frac{C_w^*}{C_w^*+1} \right) {\sigma U B^2 \ell},
    \label{eq:MHD_miya}
\end{equation}
where $C_w^* = \sigma_w (R_o^2-R_i^2)/(\sigma (R_o^2+R_i^2)) \approx 0.0575$. Thus, this prediction gives a larger pressure drop than the insulating cases, because the wall is not entirely insulating. Therefore, we consider this to be the conservative estimate. MHD drag does become more significant if eddy currents recycle through the conducting wall. This prediction is marked as the cyan line in \cref{fig:pdrop} a). Moreover, \cref{fig:pdrop} a) also shows that MHD drag is the dominating pressure sink when flow speed is $\lesssim 4$ m/s. 

Moreover, the pressure drop across an EM flowmeter and a vortex flow meter have been included in the pressure drop term $P_{diag}$. In particular, the MHD drop across the EM flowmeter is $\SI{310}{Pa}$, or $\SI{0.05}{psid}$, assuming the flow rate is $\SI{20}{GPM}$ to be conservative, the enclosed magnet is $\SI{0.1}{T}$, and the length of the uniform field is $\SI{5}{cm}$. The pressure drop in the vortex flowmeter, is estimated to be $\SI{3.93}{psid}$, or $\SI{27.03}{kPa}$ at $400^\circ$C. 

\subsection{Viscous drag}

The pressure drop due to viscous drag varies significantly when the flow is laminar or turbulent. Therefore, we can consider two different flow regimes to calculate the pressure difference due to viscous drag $\Delta P_\nu$ in a pipe. However, the exact transition from laminar to turbulence in MHD liquid metal pipe flow is strongly related to the Hartmann number, aspect ratio, and characteristic length of the pipes, and shall be determined experimentally. A critical Reynolds number $Re_c = UD/\nu$ has been reported between $\sim25$Ha \cite{zhang2017experimental} to $\sim250$Ha \cite{moresco2004experimental} at $Ha>300$, and $Re_c \lesssim 500$Ha for $Ha\lesssim1000$ \cite{hoffman1971magnetic}. With Phase I loop operating at $0\lesssim Ha \lesssim 800$, we conservatively assume the turbulent liquid metal flow under a strong magnetic field onsets near $Re_c \gtrsim 4000$. The Navier-Stokes equation gives the pressure drop via Hagen-Poiseuille equation, which applies to a laminar flow of an incompressible, Newtonian fluid through a circular pipe. The formula for the pressure drop $P_\nu$ due to viscous drag across a length $L$ of the pipe is given by:
\begin{equation}
    P_\nu = \frac{8 \mu Q L}{\pi r^4} = \frac{8 \mu U L}{r^2},
    \label{eq:HP}
\end{equation}
where $p_\nu$ is the pressure drop (in Pascals, $\text{Pa}$), $\mu$ is the dynamic viscosity of the fluid ($\mathrm{Pa \cdot s}$), $Q$ is the volumetric flow rate of the fluid (in cubic meters per second, $\mathrm{m^3/s}$), $L$ is the length of the pipe through which the fluid is flowing, $r=D/2$ is the inner radius of the pipe. This equation assumes a steady, laminar, and fully developed flow. 

This is equivalent to the Darcy-Weisbach equation at the laminar flow regime
\begin{equation}
    \Delta P_\nu = f_D \frac{L}{D}\frac{\rho U^2}{2},
    \label{eq:DW}
\end{equation}
where $f_D$ is the Darcy friction factor. For Laminar flow, $f_D = 64/Re$, \cref{eq:DW} can be written as \cref{eq:HP}. If the flow is turbulent ($Re \gtrsim 4000$), the Darcy-Weisbach equation would be more appropriate to use. Solve the Colebrook equation, 
\begin{equation}
    \frac{1}{\sqrt{f_D}} = -2 \log\left( \frac{\epsilon}{3.7 D} + \frac{2.51}{Re\sqrt{f_D}} \right) \frac{\rho U^2}{2},
    \label{eq:DW2}
\end{equation}
where $\epsilon$ is the roughness. For stainless steel, $\epsilon \approx 1.5\times 10^{-4}$ m. By solving \cref{eq:DW2}, $f_D$ can be found numerically. A smooth SS pipe has $f_D\approx 0.012$.

The static pressure at the pump inlet of a fully developed, incompressible lithium flow in a circular pipe that passes through a section of high transverse magnetic field (\cref{eq:I}) can be written as 
\begin{equation}
    P_{static} \approx P_p - \frac{\sigma U B^2 \ell}{Ha} - \frac{f_DL\rho U^2}{2d} -\frac{1}{2} \rho U^2.
    \label{eq:Ic}
\end{equation}

In addition, pipe bends will effectively increase the viscous drag of the system. The dynamic pressure is modified by the bends via a $K_b$ factor. Because the ratio between bending radius and pipe diameter $R/D\sim 2$, a $90^\circ$ bend on stainless steel, $K_b = 0.45$ \citep{carlsaw1959conduction}, the additional pressure drop is 
\begin{equation}
    \Delta P_{bend} = N K_b P_d = N K_b \frac{\rho U^2}{2}
    \label{eq:bends}
\end{equation}
For an $1$ L/s ($\SI{15.85}{GPM}$) flow in the Phase I system with six bends, the largest estimated pressure drop due to the bends is $\sim 2.7 P_d \approx 0.81~$psi. Therefore, this is relatively small compared to MHD and viscous drag in the system.  

\subsection{Heating and cooling}

\begin{figure}[t]
    \centering
    \includegraphics[width=1\linewidth]{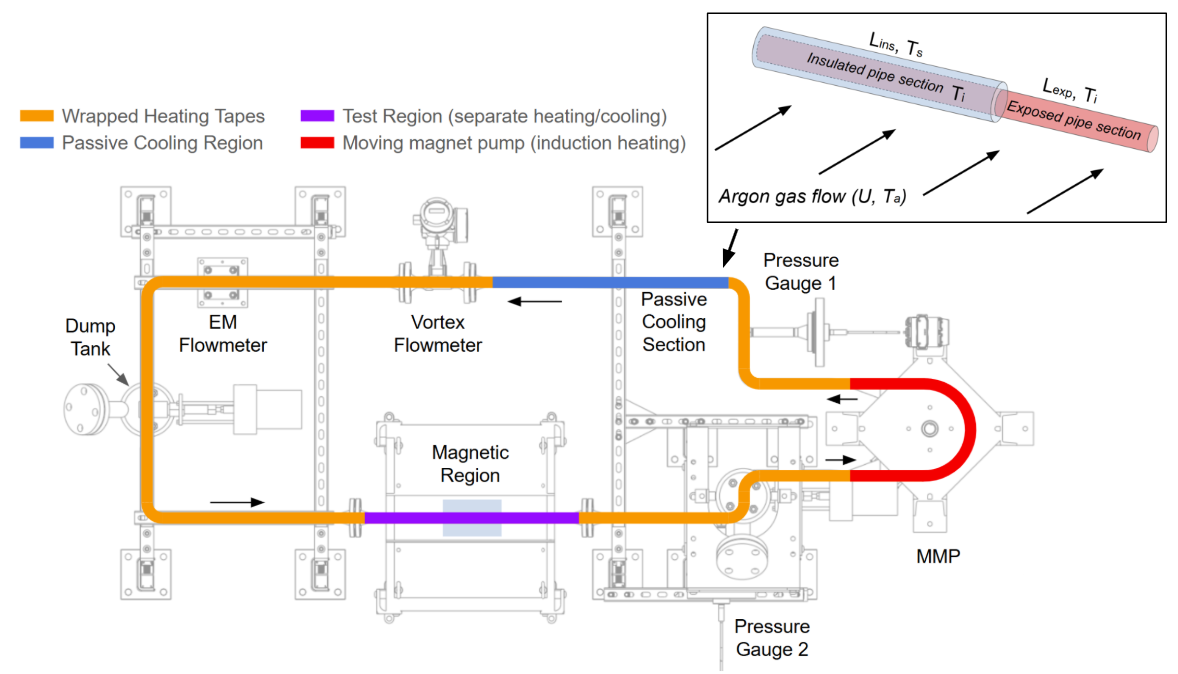}
    \caption{Top view of the thermal sections in the lithium main loop and schematics of the simplified 1D model for heat transfer across the heated piping with insulated and exposed sections, respectively.}
    \label{fig:heating}
\end{figure}
\begin{figure}
    \centering
    \includegraphics[width=0.7\linewidth]{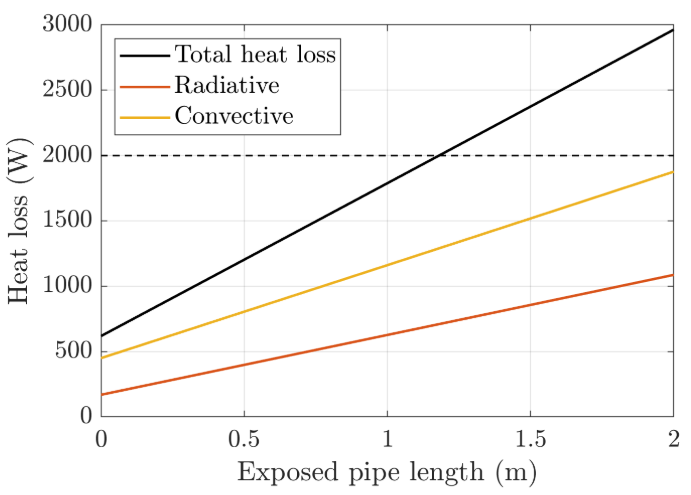}
    \caption{Heat loss as a function of exposed pipe length at air circulation speed of $U = \SI{3}{m/s}$. }
    \label{fig:heatloss}
\end{figure}
\Cref{fig:heating} shows a top-view drawing of the loop system, with the main heating/cooling sections highlighted in colors. The orange color represents normal piping sections wrapped with heating tape. The red curve shows the piping section under the MMP induction heating. The passive cooling section shown in blue (length varies) does not have insulation and is exposed to the Argon environment inside the glovebox. The purple section represents the test piping region inside the magnetic field, and is independently heated and cooled if necessary. 

For the analysis, the loop system can be treated as two piping sections in series: a constant pipe surface temperature at $T_i = 400^\circ$C ($\SI{673.15}{K}$) from the PID-controlled heating tape, next to an insulation-free exposed pipe section that is also wrapped by the PID-controlled heating tape at $T_i$. The ambient argon gas temperature $T_a = \SI{26.85}{^\circ C}$ ($\SI{300}{K}$) is assumed to be the temperature at the outer layer of the insulation. The schematic for this simplified 1D system is shown in \cref{fig:heating}. 

Because there is no active heating at the test region for the Phase I loop, and the gloveroom has a built-in HVAC cooling system capable of handling $\SI{6.59}{kW}$ heating power, we can utilize the internal circulation of argon gas to cool the pipe. Specifically, \cref{fig:heating} shows a passive cooling section between the vortex flowmeter and the pressure gauge. 

\Cref{fig:heatloss} shows total heat loss and its radiative and convective components as a function of insulation-free exposed pipe length. The argon flow speed is in the range of $1-\SI{5}{m/s}$, which is a normal flow speed range of a commercial box fan. A total of approximately $\SI{2}{kW}$ heat is generated from the loop to run the moving magnet pump at top speed while keeping lithium at a maximum temperature of $\SI{400}{^\circ C}$. \Cref{fig:heatloss} demonstrates that an $\SI{1.2}{m}$-long exposed pipe is needed to exhaust the generated heat from the loop with a $\SI{3}{m/s}$ argon gas flow.

\bibliographystyle{elsarticle-num} 
\bibliography{ref}





\end{document}